\documentclass[9pt]{costum}

\usepackage{lipsum} 
\usepackage[version=4]{mhchem}
\usepackage{siunitx}
\DeclareSIUnit\Molar{M}

\title{Intrinsic timescales of spiking activity in humans during wakefulness and sleep}
\author[1]{Annika Hagemann} 
\author[2]{Marcel Stephan Kehl}
\author[1]{Jonas Dehning}
\author[1]{F.~Paul Spitzner}
\author[2]{Johannes Niediek}
\author[4]{Michael Wibral}
\author[2]{Florian Mormann}
\author[1,5,6*]{Viola Priesemann}
\affil[1]{Max Planck Institute for Dynamics and Self-Organization, Göttingen, Germany}
\affil[2]{Department of Epileptology, University of Bonn Medical Centre, Bonn, Germany}
\affil[4]{Campus Institute for Dynamics of Biological Networks, Georg-August University, Göttingen, Germany}
\affil[5]{Institute for the Dynamics of Complex Systems, University of Göttingen, Göttingen, Germany}
\affil[6]{Leibniz ScienceCampus Primate Cognition, Göttingen, Germany}

\corr{viola.priesemann@ds.mpg.de}{Viola Priesemann}

\begin{document}

\maketitle

\begin{abstract}
Information processing in the brain requires integration of information over time. 
Such an integration can be achieved if signals are maintained in the network activity for the required period, as quantified by the intrinsic timescale. While short timescales are considered beneficial for fast responses to stimuli, long timescales facilitate information storage and integration. 
We quantified intrinsic timescales from spiking activity in the medial temporal lobe of humans. We found extended and highly diverse timescales ranging from tens to hundreds of milliseconds, though with no evidence for differences between subareas.
Notably, however, timescales differed between sleep stages and were longest during slow wave sleep. This supports the hypothesis that intrinsic timescales are a central mechanism to tune networks to the requirements of different tasks and cognitive states.
\end{abstract}

\section*{Introduction}
\vspace{0.5cm}
When processing language, navigating in space or performing actions, the brain must be able to integrate information over time. An illustrative example is language processing, as the beginning of a sentence must remain accessible in order to correctly understand its end. One mechanism to achieve such an integration is by means of the network activity. If a signal reverberates in the network activity for the required period of time, it remains accessible for later retrieval~\cite{jaeger2001echo, cramer2020control}.
A direct approach to quantify how long a signal is maintained in the network activity has recently been introduced~\cite{murray_hierarchy_2014, wilting_inferring_2018}. 
It measures the intrinsic timescale $\tau$ of a brain area as the decay in the autocorrelation function of the local spiking activity. This decay, although measured from only a tiny subset of neurons, is nonetheless representative for the local circuit~\cite{wilting_between_2019, wilting_operating_2018}.
While long $\tau$ imply that a signal persists in the system for a long period of time, short $\tau$ imply less dependence on previous activity. 
\par

From a theoretical perspective, different intrinsic timescales are optimal for different functions. While short intrinsic timescales $\tau$ are considered beneficial for fast responses to stimuli and fast forgetting, long $\tau$ facilitate information storage and integration~\cite{cramer2020control, wilting_operating_2018,cavanagh_diversity_2020, golesorkhi_brain_2021}.
Thus, it has been hypothesized that, depending on functional requirements, the intrinsic timescale varies both \emph{spatially} (depending on the brain area)~\cite{murray_hierarchy_2014, wilting_inferring_2018,hasson2008hierarchy,hasson2015hierarchical, chaudhuri_large-scale_2015, gao2020neuronal, shafiei_topographic_2020}, and \emph{temporally} (depending on the current task)~\cite{meisel2017interplay,cirillo2018neural,wasmuht_intrinsic_2018,gao2020neuronal}.
These hypotheses have been investigated experimentally in different species, but results from spiking activity in humans are missing so far. \par

A \emph{spatial} organization of intrinsic timescales along the sensory processing hierarchy has been suggested conceptually and found experimentally for several mammalian species~\cite{murray_hierarchy_2014, wilting_operating_2018,hasson2008hierarchy,hasson2015hierarchical, siegle_survey_2019, chaudhuri_large-scale_2015, gao2020neuronal}:
For example, spike recordings from macaque suggest a hierarchy of intrinsic timescales that is directly related to the functional hierarchy of information processing, with shorter timescales in sensory areas and longer timescales in prefrontal areas~\cite{murray_hierarchy_2014}. In line with this finding, intrinsic timescales from spike recordings from the mouse visual system were found to increase from thalamus to early and then higher cortical visual areas, suggesting a correlation with the anatomical hierarchy~\cite{rudelt2022mouse}.
Together, these studies suggest that early or unimodal areas show short $\tau$ for fast processing, whereas higher, multimodal cortical areas show longer $\tau$ to facilitate integration of information~\cite{hasson2008hierarchy, murray_hierarchy_2014, chaudhuri_large-scale_2015, hasson2015hierarchical, gao2020neuronal}. \par

A \emph{temporal} variation of timescales is supported by spike recordings from different mammals during task conditions, as well as during different vigilance states.
Recordings from mouse frontal and parietal cortex revealed prolonged intrinsic timescales during REM sleep and shorter timescales during non-REM sleep~\cite{meisel2017interplay}. 
Furthermore, evidence from dorsal premotor cortex~\cite{cirillo2018neural} and lateral prefrontal cortex~\cite{wasmuht_intrinsic_2018} of rhesus macaques during memory tasks suggest that the intrinsic timescale of individual neurons is related to the neuron's involvement in the task. In dorsal premotor cortex, neurons with longer timescales showed an increased spatial response coding in the delay period~\cite{cirillo2018neural}. In lateral prefrontal cortex, neurons with short timescales were more involved at an early stage during memory encoding, while long-timescale neurons were more involved at a later stage, dominating coding in the delay period~\cite{wasmuht_intrinsic_2018}.
Thus, it has been hypothesized that the brain is able to \emph{tune} network timescales depending on the current functional requirements~\cite{wilting_operating_2018, cramer2020control, cavanagh_diversity_2020, golesorkhi_brain_2021}.\par

Despite vast evidence from spiking activity in non-human mammals, obtaining reliable estimates of $\tau$ for humans is hindered by the invasive procedure when recording spiking activity. Therefore, intrinsic timescales, or related measures, have only been inferred from fMRI~\cite{watanabe_atypical_2019}, EEG~\cite{watanabe_atypical_2019,zilio2021intrinsic} and ECoG data~\cite{honey2012slow,gao2020neuronal} in the past. Intrinsic timescales from fMRI data revealed longer timescales in frontal and parietal cortices and shorter timescales in sensorimotor, visual, and auditory areas~\cite{watanabe_atypical_2019}. Furthermore, intrinsic timescales differed between autism patients and healthy individuals~\cite{watanabe_atypical_2019} and timescales obtained from ECoG data revealed differences between sensory regions and higher order regions~\cite{honey2012slow}. 
Most recently, a hierarchy of intrinsic timescales along the anatomical hierarchy from sensory to association regions was confirmed in human ECoG data~\cite{gao2020neuronal}. This comprehensive study further demonstrated a reduction of timescales with the age of the subject, and an increasing timescale in prefrontal cortex during working memory tasks.
Lastly, human EEG data~\cite{zilio2021intrinsic} showed differences in timescale-related measures depending on the vigilance state, suggesting that timescales are prolonged during sleep, with longest timescales during sleep stage N3.
\par

While these findings from humans are qualitatively in line with findings from other mammalian species, the specific values of intrinsic timescales in humans are an order of magnitude smaller than the intrinsic timescales estimated from spiking activity of other mammals~\cite{murray_hierarchy_2014, wilting_operating_2018, wasmuht_intrinsic_2018, cirillo2018neural}. 
This may be caused by the fact that human timescales were obtained from coarse data, and thus may not reflect the timescale of local spiking activity. Although~\cite{gao2020neuronal} found a significant correlation between timescales from spiking data and timescales from ECoG in macaque recordings, it remains to be assessed precisely to which extent these coarse values do indeed reflect the intrinsic timescale of local spiking activity.\par

Here, we quantified for the first time intrinsic timescales of spiking activity in humans. Using recordings from the medial temporal lobe of subjects with medically intractable focal epilepsy, we addressed three main questions:
\begin{enumerate}
    \item How long are intrinsic timescales of spiking activity in humans and how do they compare to other species?
    \item Do intrinsic timescales in humans change depending on the cognitive state, more specifically, depending on the vigilance state?
    \item Do timescales in humans differ across different subregions of medial temporal lobe (MTL)?
\end{enumerate}

To address these questions, we measured intrinsic timescales in spiking activity from 24 subjects with medically intractable focal epilepsy during rest.
Recordings spanned different subregions of MTL, with depth electrodes placed in hippocampus, amygdala, parahippocampal cortex and entorhinal cortex.
In addition, we recorded spiking data of six subjects during sleep, enabling a comparison of timescales across sleep stages including REM, N1, N2 and N3.
Finally, we put our results into relation of previous analyses of timescales in different mammalian species and across brain areas.

\begin{figure}[t!]
    \centering
    \includegraphics[width=0.999\textwidth]{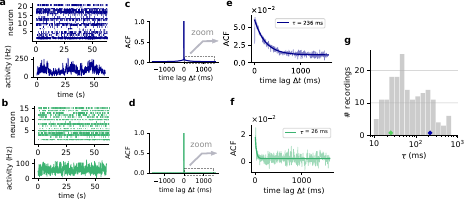}
    \caption{\footnotesize{Spontaneous spiking activity in human MTL shows extended and highly diverse intrinsic timescales $\tau$. \textbf{a}, \textbf{b} Example raster plots of spike times in human MTL and the corresponding population activities $A_t$ obtained by summing spikes across all single units to 4~ms time bins. Plots show 60~s segments of two different recordings in MTL. \textbf{c},\textbf{d} Autocorrelation functions (ACFs) of population activity $A_t$. Across recordings, the ACF drops steeply from delay $\Delta t=0$ to $\Delta t \neq 0$, but the "floor", i.e. the ACF for $\Delta t > 0$, clearly decays exponentially (see magnified view in \textbf{e,f}). This exponential decay defines the intrinsic timescale $\tau$. \textbf{e,f} ACFs and fitted exponentials of the same example recordings. The two recordings have clearly different timescales $\hat{\tau}_1 = 236$~ms and $\hat{\tau}_2 = 26$~ms. 
    \textbf{g} Distribution of intrinsic timescales $\hat{\tau}$ across recordings and subjects. Timescales are highly diverse, ranging from $\sim$10~ms to $\sim$700~ms (median 54~ms). See Figs.~\ref{suppfig:ACFs_first}-\ref{suppfig:ACFs_last} for individual fits.}}
    \label{Fig1}
\end{figure}

\section{Results}
\vspace{0.5cm}

To quantify the intrinsic timescale $\tau$, we estimate the autocorrelation function (ACF) of the population activity $A_t$, where $A_t$ is calculated from time-binned spike times of either all recorded neurons in a hemisphere, or the neurons of a specific MTL subregion (Fig.~\ref{Fig1}).
For activity that features even slight temporal correlations (such as 
commonly found in non-human mammalian species \cite{murray_hierarchy_2014, wilting_inferring_2018, wilting_between_2019}) the ACF can decay exponentially, where the amplitude of the ACF reflects the degree to which the system was sampled~\cite{wilting_inferring_2018}, and the time constant of the decay is a subsampling-invariant estimate of the intrinsic timescale $\tau$ (Fig.~\ref{Fig1}c, d)~\cite{wilting_inferring_2018, murray_hierarchy_2014}.

We find a broad distribution of intrinsic timescales (Fig.\ref{Fig1}\textbf{g}), with $\tau$ in the range from 10~ms to 700~ms (median $\bar{\tau}_\textrm{human}=54$~ms) and variability both, across and within subjects (see also Figs.~\ref{suppfig:ACFs_first}-\ref{suppfig:ACFs_last}).
Variability \emph{between} subjects may reflect the precise position of recording electrodes, or subject-specific differences including age, gender, or medication. Variability \emph{within} subjects over time may reflect different cognitive states at the time of recording, or small changes in the position of recording electrodes. Overall, our results suggest that intrinsic timescales in human MTL are not restricted to a single value, but range from tens to hundreds of milliseconds. 
\par 

\begin{figure}[h!]
    \centering
    \includegraphics[width=0.99\textwidth]{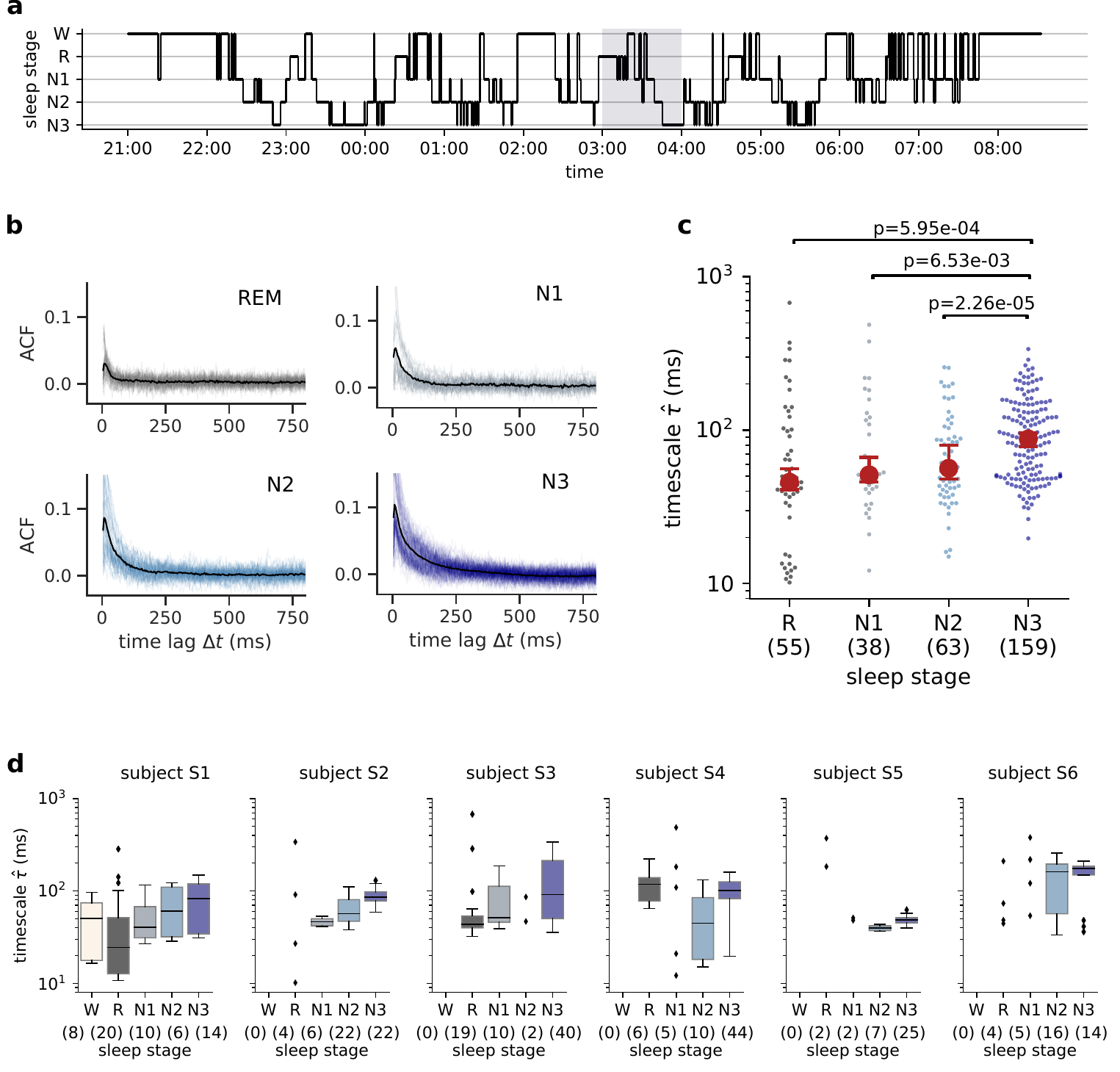}
    \caption{\footnotesize{Intrinsic timescales $\hat{\tau}$ are longest during slow wave sleep. \textbf{a} Example trace of recorded sleep stages of one subject. For each of the six subjects, we analyzed one hour of sleep. \textbf{b} All autocorrelation functions (ACF) of spiking activity in different sleep stages (REM, N1, N2, N3). While ACFs during REM sleep tend to decay rapidly, ACFs during sleep stage N3 exhibit a slower decay, i.e. spiking activity is correlated across a longer period of time. Black lines show averages across all ACFs. 
    \textbf{c} Intrinsic timescales $\hat{\tau}$ differ between sleep stages (mixed effect ANOVA with subjectID as random effect, $p < 10^{-4}$). In particular, activity in sleep stage N3 (slow wave sleep) shows longer intrinsic timescales than activity in sleep stages N2, N1 and REM. Annotated values show uncorrected p-values for pairwise comparisons (pairwise mixed effect ANOVA with subjectID as random effect, $\alpha = 0.05/6 \approx 8\cdot 10^{-3}$ required after Bonferroni correction). Numbers in brackets indicate the number of 90 seconds recording segments in the respective sleep stage. Median and 95\% confidence intervals are annotated in red.
    \textbf{d} Results displayed for each individual patient. For less than 6 recording segments, we only show the observed $\tau$ as dots, not the distribution.}}
    \label{Fig2}
\end{figure}

To investigate whether the brain temporarily tunes the network timescale to processing requirements, we measured $\tau$ across vigilance states (Fig.~\ref{Fig2}). In total, we analyzed 6 hours of sleep recordings from $n=6$ patients (dataset 2, see table~\ref{table_sleep}). Recordings were classified into rapid eye movement sleep (REM), and sleep stages N1, N2 and N3, and $\tau$ was estimated on 90~s recording segments. We found significant differences across $\tau$ from different vigilance states (mixed effect ANOVA on the log-transformed timescales with subjectID as random effect, $F=7.5$, $p < 10^{-4}$). In particular, activity during slow wave sleep (N3) showed longer $\tau$ than activity during REM, N1 and N2 ($p\approx 6\cdot10^{-4}$, $p\approx 7\cdot10^{-3}$, $p\approx 2\cdot10^{-5}$, respectively, $\alpha = 0.05/6 \approx 8\cdot 10^{-3}$ required after Bonferroni correction). 
This finding suggests that $\tau$ is not a static network property, but can vary depending on the vigilance state. 

\par
\begin{figure}
    \centering
    \includegraphics[width=0.8\textwidth]{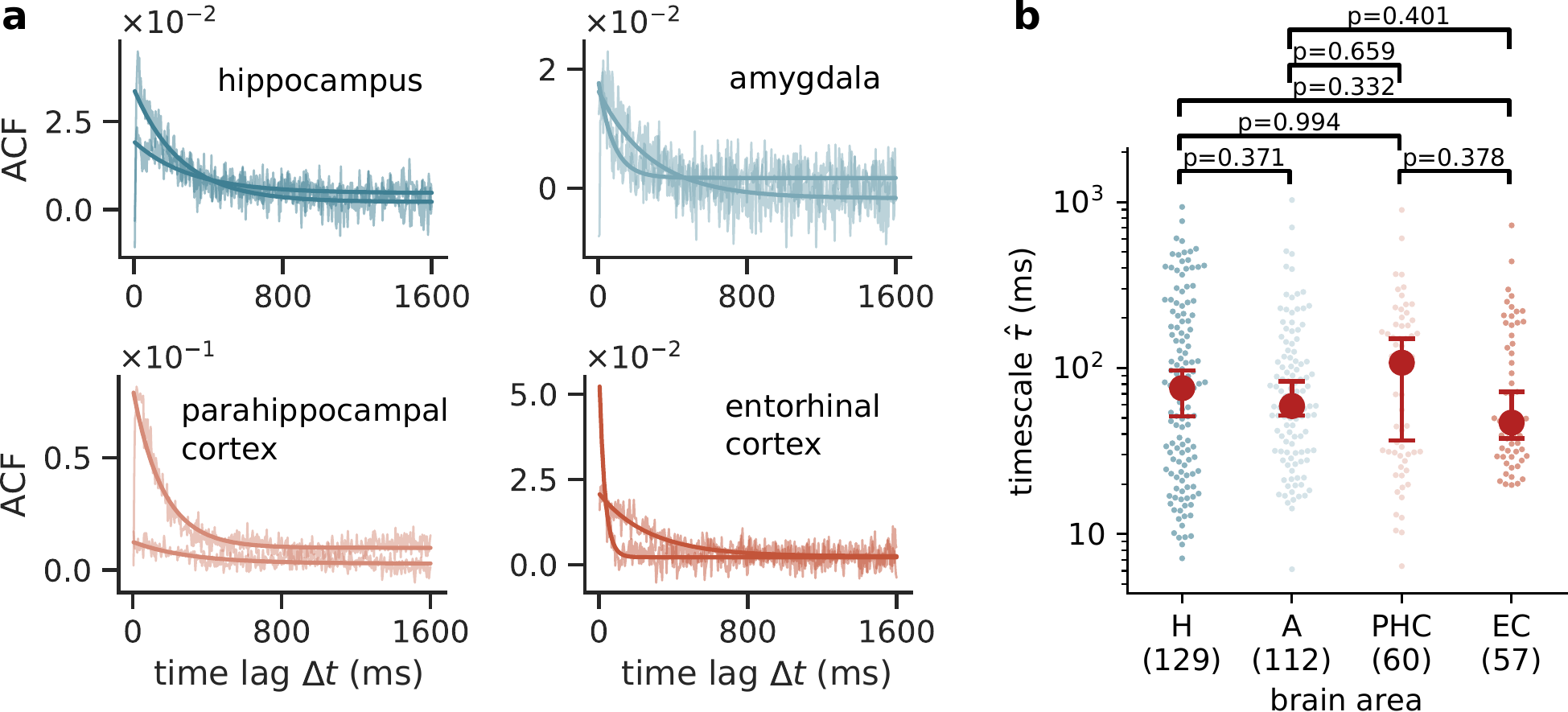}
    \caption{\footnotesize{No significant difference between intrinsic timescales $\hat{\tau}$ across subregions of human medial temporal lobe (MTL). 
    \textbf{a} Examples of autocorrelation functions (ACFs) of spiking activity in hippocampus (H), amygdala (A), parahippocampal cortex (PHC) and entorhinal cortex (EC). The ACF typically decays exponentially, but $\hat{\tau}$ is highly diverse within each subregion. The plot shows two exemplary ACFs per subregion. For all ACFs, see Fig.~\ref{suppfig:autocorrelations_regions}. 
    \textbf{b} The intrinsic timescales $\hat{\tau}$ do not differ significantly across subregions (mixed effect ANOVA on the log-transformed timescales with subjectID as random effect, $F=0.62$, $p \approx 0.60$). Numbers in brackets indicate number of recordings in the respective subregion. Median and 95\% confidence intervals are annotated in red.}}
    \label{Fig3}
\end{figure}

To investigate whether different subregions of the MTL operate at different timescales, we estimated $\tau$ separately for amygdala, entorhinal cortex, parahippocampal cortex and hippocampus (Fig.~\ref{Fig3} dataset 1). We consistently found exponentially decaying ACFs in all subregions, but between subregions $\tau$ did not differ significantly (mixed effect ANOVA on the log-transformed timescales with subjectID as random effect, $F=0.62$, $p \approx 0.60$).
However, $\tau$ was highly diverse in each subregion. Thus, if there was a systematic difference between subregions of MTL, the effect size was not sufficiently large to detect it given the overall high variability.
\par

\begin{figure}[h!]
    \centering
    \includegraphics[width=0.6\textwidth]{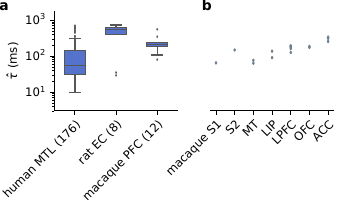}
    \caption{\footnotesize{Intrinsic timescales across different mammals. \textbf{a} Using our analysis methodology, $\hat{\tau}$ estimated for human MTL during rest, rat hippocampus (H) during foraging, and macaque prefrontal cortex (PFC) during a working memory task. Numbers in brackets indicate the respective number of analyzed recordings and Figs.~\ref{suppfig:autocorrelations_regions} and \ref{suppfig:autocorrelations_species} show the individual fits.
    \textbf{b} Timescales of various cortical areas of macaque \cite{murray_hierarchy_2014}, measured and reported by Murray et al.~\cite{murray_hierarchy_2014}.}}
    \label{Fig4}
\end{figure}

To relate the measured timescales in humans to previous results from other mammalian species, we repeated our analysis on previously published spiking data (Fig.~\ref{Fig4}a) and compared with published estimates of $\tau$, also from spiking activity (Fig.~\ref{Fig4}b).
More specifically, we evaluated spiking activity of macaque prefrontal cortex (PFC) during a memory task~\cite{pipa_performance-_2009}
and of rat hippocampus (H) during foraging \cite{rat_EC_data, mizuseki_theta_2009} using our methods, and for comparison replotted the estimates of $\tau$ from different cortical areas in macaque~\cite{murray_hierarchy_2014}.
All mammalian timescales were between tens and hundreds of milliseconds, similar to the range we estimated in humans
(Fig.~\ref{Fig4}a,b).
Interestingly, we found that both, for macaque PFC and rat EC $\tau$ was longer than in human MTL ($\bar{\tau}_{macaque}=215$~ms, $p\approx 0.0002$, and $\bar{\tau}_{rat}=553$~ms, $p\approx 0.004$; Mann Whitney-U test)
and also differed between human EC and rat EC (median $\bar{\tau}_\textrm{EC, human}=47$~ms, and $\bar{\tau}_\textrm{EC, rat}=553$~ms; $p\approx 0.004$, Mann Whitney-U test). Although the different recording conditions hinder a rigorous quantitative comparison of the species, the results indicate that intrinsic timescales can not only differ depending on brain area and task, but may also vary between species.


\section*{Discussion}
\vspace{0.5cm}
We found that spiking activity in human MTL shows extended and highly diverse intrinsic timescales, ranging from tens to hundreds of milliseconds. Across patients and cortical subregions, the timescales could be extracted reliably from the autocorrelation functions of spiking activity, because they were widely consistent with the expected exponential decay. Finding intrinsic timescales in this range is in line with results on various mammalian species~\cite{murray_hierarchy_2014, wasmuht_intrinsic_2018, wilting_between_2019, cirillo2018neural}, suggesting a functional advantage of maintaining signals in the network activity for extended periods of time.\par

Intrinsic timescales during slow wave sleep were longer than during other vigilance states. This supports the notion that intrinsic timescales are not a static network property, but can vary depending on the cognitive state or task requirements. 
Our finding is consistent with that from earlier analyses on human LFP recordings in MTL~\cite{priesemann_neuronal_2013}. In that study, neuronal avalanches were found to be larger in SWS than in REM, indicating longer intrinsic timescales~\cite{wilting_between_2019}. Similarly, longer timescales have been also found in human EEG recordings during sleep compared to the awake state \cite{zilio2021intrinsic}.
These longer timescales and larger avalanches may facilitate integration of information across brain areas and hence support memory consolidation in deep sleep~\cite{rasch_about_2013}.
However, our findings differs from a previous study that found a disruption of long timescales during non-REM sleep (N1, N2, N3) in rat frontal and parietal cortex~\cite{meisel2017interplay}. 
A possible reason is that the recordings were obtained from different species,  as well as different brain areas.

While our results support the notion of a \emph{state-dependent} tuning of intrinsic timescales, we found no evidence for a \emph{spatial} difference across subregions of MTL. This could be explained by the fact that all analyzed subregions were within MTL, high in the hierarchy of information processing. Systematic differences, if present, are therefore likely to be smaller than the differences between sensory areas (low-level) and prefrontal areas (high-level), which have been observed in previous studies~\cite{murray_hierarchy_2014, hasson2008hierarchy}. Therefore, our results do not contradict the existing studies that suggest a hierarchy of timescales. 
\par

Compared to macaque prefrontal cortex and rat hippocampus, timescales in human MTL were comparatively short (medians $\bar{\tau}_{\text{human}}=54$~ms, $\bar{\tau}_{\text{macaque}}=215$~ms, $\bar{\tau}_{\text{rat}}=553$~ms). This might reflect a general difference between the species, but it might also be related to different recording conditions and cognitive states. The comparatively short timescales in humans may e.g. reflect that they were recorded at rest, whereas most of the recordings in other species were performed in a task context (memory  or foraging tasks). Moreover, human recordings were obtained from epilepsy patients, who typically receive anti-epileptic drugs. In most cases, these drugs reduce the excitability of neurons and might induce more subcritical dynamics, corresponding to shorter intrinsic timescales \cite{meisel2015intrinsic, meisel_antiepileptic_2019}. \par

\begin{figure*}
    \includegraphics[width=0.99\textwidth]{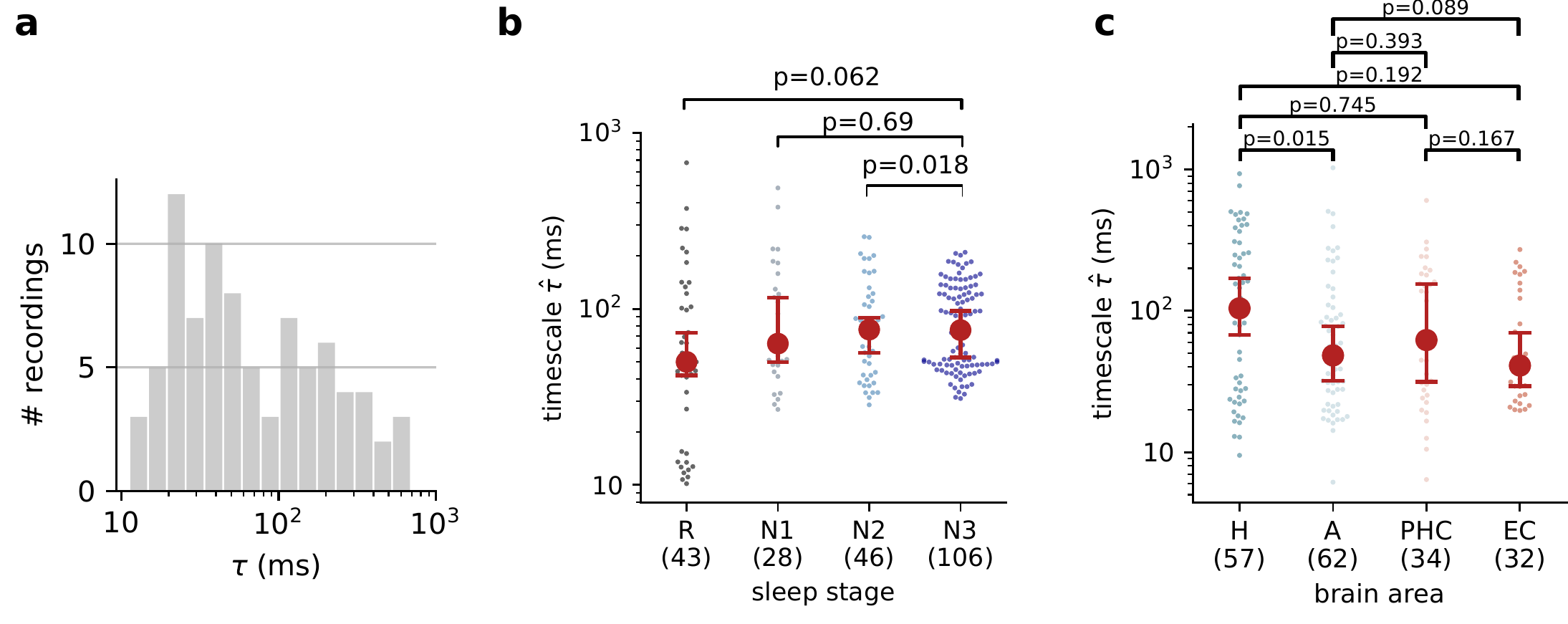}
    \caption{Main analyses excluding the data obtained from the ipsilateral hemispheres of the subjects. \textbf{a} Distribution of estimated intrinsic timescales in human MTL. Median $\bar{\tau}'_\textrm{human}=49$~ms, compared to $\bar{\tau}_\textrm{human}=54$~ms when including all recordings. \textbf{b} Comparison of intrinsic timescales across sleep stages. Compared to the overall finding (Fig.~\ref{Fig2}), the difference between sleep stages is reduced in this subset of the data. \textbf{c} Comparison of intrinsic timescales across subregions of MTL (overall finding in Fig.~\ref{Fig3}). Annotated values in \textbf{b}, \textbf{c} show uncorrected p-values for pairwise comparisons (pairwise mixed effect ANOVA with subjectID as random effect, $\alpha = 0.05/6 \approx 8\cdot 10^{-3}$ required after Bonferroni correction).}
    \label{suppfig:only_contra}
\end{figure*}

Although we evaluated seizure-free activity only, dynamics were likely affected by medication and epilepsy, which could limit the generalizability of our findings to healthy humans. Excluding data from the ipsilateral hemisphere of patients with identified seizure onset zone reduces the observed differences between sleep stages (see Fig~\ref{suppfig:only_contra}). Although this may be caused by the mere reduction in recording numbers, and the resulting higher uncertainty of ensemble averages, it cannot be excluded that the increased timescale during N3 is related to or caused by epilepsy.

In principle, estimating the intrinsic timescale $\tau$ from spiking activity is straight-forward, but the technical details across different studies differ.
Our analysis relied on the population activity, measuring the autocorrelation function of all spikes that were recorded in a specific brain area. Alternatively, one can separately evaluate the ACF of each single neuron \cite{wasmuht_intrinsic_2018, cirillo2018neural} and subsequently average over all neurons to obtain an estimate for the population~\cite{murray_hierarchy_2014, cirillo2018neural}. Furthermore, time can be discretized (binned) at around 4~ms~\cite{wilting_between_2019}, reflecting the time of causal spike propagation, or at around 50~ms~\cite{murray_hierarchy_2014, cirillo2018neural}. Likewise, the intrinsic timescales obtained from coarse-sampled-data, like LFP, EEG or BOLD signals, might not be directly comparable (e.g.~\cite{gao2020neuronal}) because of the additional filtering and spatial averaging. Such coarse-graining is known to heavily affect other dynamical measures \cite{neto_unified_2020}.
All these choices may influence the absolute values of the intrinsic timescale that are obtained.

The diverse intrinsic timescales we observed might reflect various mechanisms. They might reflect cellular or synaptic dynamics of individual neurons, the statistics of the external input, or they may arise as an emergent network property~\cite{huang_once_2017}.
However, from observations alone, these causes are difficult to distinguish.
For example, long timescales can be generated on the single-neuron level by spike-rate adaptation or facilitation, mediated e.g. by the slow dynamics of calcium and the vesicle cycle.
Alternatively, on the level of the input, signals from the external world clearly show temporal structure (including long timescales), which might be reflected in the measured timescale.
On the network level, recurrent excitatory connections are decisive for signals to reverberate in the network, giving rise to long intrinsic timescales~\cite{wilting_between_2019,cramer2020control}. This mechanism is harnessed in artificial networks for information integration~\cite{maass2002real,jaeger2001echo,loidolt_sequence_2020,cramer2020control}.
Finding such long intrinsic timescales in high cortical areas, far from the sensory input, could indicate that they are not actively cancelled across the hierarchy, but are maintained or even prolonged, which stresses their role for information processing.

Overall, these first measurements of intrinsic timescales of spiking activity in humans support the notion that intrinsic timescales are not a static, homogeneous network property. They are highly diverse within cortical areas, and apparently change with sleep stages, demonstrating that timescales can vary systematically depending on the cognitive state.

\section*{Methods}

\subsubsection*{Acquisition and pre-processing of intracranial recordings}
We analyzed intracranial recordings from $n=24$ patients with medically intractable focal epilepsy\footnote{The dataset is partially equal to the dataset used in~\cite{hagemann2021assessing}. Therefore, the methods sections partially overlap.}. The data was recorded at the Department of Epileptology at the University of Bonn Medical Center. For pre-surgical evaluation, patients were implanted with depth electrodes in different regions of the medial temporal lobe, including hippocampus (H), amygdala (A), parahippocampal cortex (PHC) and entorhinal cortex (EC). Three electrodes were placed in hippocampus and one electrode in each of the other subregions. The location of microwires was verified using post-implantational CT scans co-registered to pre-implantational MRI scans. All wire bundles were confirmed to be located in the designated target regions in each patient.
Each electrode contained 8 microwires, with which single-unit recordings could be performed. All patients had given written informed consent to participate in this study, which was approved by the Medical Institutional Review Board in Bonn. Recordings were performed continuously for pre-surgical monitoring for a typical duration of 7-14 days. 
Data was sampled at 32 kHz and filtered between 0.1 and 9000 Hz. Spike sorting was performed using the Combinato package~\cite{niediek_reliable_2016}, using the standard parameters proposed by the authors. Sorted units were classified manually as single units, multi-units, or artifacts, using the Combinato GUI~\cite{niediek_reliable_2016}. The main classification criterion for putative units was the signal's shape, but also the amplitude and distribution of inter-spike-intervals, which allows excluding artifacts due to e.g. supply voltage. For further analysis, we only used spikes of identified single units and excluded artifacts and multi-units. Thereby, we minimized the number of artifacts, which can potentially impact subsequent analyses (multi-units typically contain more artifacts than identified single units). We analyzed two different datasets:\\

\textbf{Dataset 1: Awake 10-minute recordings from $n=20$ patients.}
For each of the patients, we analyzed one 10-minute reference recording, obtained in a seizure-free interval after the surgery, as well as several pre-ictal recordings, spanning 10 minutes prior to seizure onset. 
Table~\ref{tab:rec_overview} summarizes the analyzed patients and recordings. \\

\textbf{Dataset 2: 1-hour sleep recordings from $n=6$ patients.}
We analyzed a total of 6 hours sleep from $n=6$ patients. For each patient, one hour of sleep was recorded and the sleep stage was classified to rapid eye movement sleep (REM), and sleep stages N1, N2, N3. Sleep staging was performed for each patient on the entire night of sleep by an experienced rater based on surface EEG (C3, C4, Cb1, Cb2, F3, F4, O1, O2), EOG as well as chin EMG.
Table \ref{table_sleep} summarizes the analyzed patients and recordings.

\subsubsection*{Definition of the activity $A_t$}
The activity $A_t$ is defined as the number of active neurons in discrete time bins $\delta t$. Implanted depth electrodes can, however, only record a tiny fraction of all neurons, and hence one only observes a subset of the activity $A_t$. In general, such spatial subsampling can lead to strong biases in inferred system properties~\cite{priesemann2009subsampling,priesemann_spike_2014,levina_subsampling_2017,Wilting2019_25}.
The intrinsic timescale $\tau$, however, is invariant under spatial subsampling~\cite{wilting_inferring_2018}. \\
In the following, we thus also denote the \textit{sampled} activity by $A_t$. It is defined as the number of \textit{sampled} active neurons at time $t$. To obtain $A_t$ from recorded spike times, all spikes recorded in one brain area are pooled and binned to $\delta t = 4$~ms time bins. The time step $\delta t$ was chosen to reflect the typical propagation time of spikes between neurons.

\subsubsection*{Definition of the autocorrelation}
Given the recorded activity $A_t$, we define the autocorrelation at time lag $\Delta t$ as
\begin{align}
    \textstyle C(\Delta t) &=\frac{\textrm{Cov}[A_t,A_{t+\Delta t}]}{\textrm{Var}[A_t] } \nonumber \\
     \textstyle &= \frac{\sum_{t=0}^{T-\Delta t} (A_t-\bar{A}_t)(A_{t+\Delta t}-\bar{A}_{t+\Delta t})}{\sum_{t=0}^{T-\Delta t} (A_t - \bar{A}_t)^2},
     \label{eq:ACF}
\end{align}
where $\bar{A}_t$ and $\bar{A}_{t+\Delta t}$ denote the mean activity of the original and the delayed time series, respectively, and $T$ the duration of the recording. This definition of the autocorrelation function $C(\Delta t)$ is equivalent to the standard definition of the Pearson correlation coefficient $\rho_{A_t, A_{t+\Delta t}} =\frac{\textrm{Cov}[A_t,A_{t+\Delta t}]}{\sigma_{A_t}\sigma_{A_{t+\Delta t}}  }$, with standard deviations $\sigma_{A_t}$, $\sigma_{A_{t+\Delta t}}$, as long as $\{A_t\}_{t=0}^T$ is a stationary process and thereby $\sigma_{A_{t}}=\sigma_{A_{t+\Delta t}} $.

\subsubsection*{Estimation of the intrinsic timescale}
The intrinsic timescale $\tau$ is defined as the decay time of the autocorrelation function $C(\Delta t)$ of spiking activity. It can therefore be estimated by fitting an exponential decay 
\begin{align}
f(\Delta t)=A \exp{(-\Delta t/\hat{\tau})}+B
\end{align}
to the measured autocorrelation $C(\Delta t)$. The additional offset $B$ in the fit function $f(\Delta t)$ accounts for contributions with long timescales that do not decay substantially within the recording time, and it compensates for small non-stationarities in $A_t$~\cite{murray_hierarchy_2014}.\\
In total, given the activity $A_t$, estimation is performed in two steps \cite{wilting_inferring_2018}:
\begin{enumerate}
    \item Compute the autocorrelation function $C(\Delta t)$ for different time delays $\Delta t$ (equation \ref{eq:ACF}).
    \item Fit an exponential decay $f(\Delta t)=A \exp{(-\Delta t/\hat{\tau})}+B$ to the autocorrelation function to obtain an estimate for the intrinsic timescale $\tau$.
\end{enumerate}
All analyses were performed using the Python toolbox of the MR estimator \cite{spitzner2021mr}. For human recordings, we used time delays $\Delta t \in [4 \text{ ms}, 1600 \text{ ms}]$, which is on the order of several autocorrelation times of our data.
Confidence intervals of estimates for single recordings were obtained via a block bootstrap procedure: recordings were divided into segments of 20~s length and estimation was performed on random subsets of segments (see \emph{stationarymean} method in~\cite{spitzner2021mr}).

\subsubsection*{Exclusion criteria for intrinsic timescale estimation}
For the reliable estimation of the intrinsic timescale, the autocorrelation $C(\Delta t)$ must be consistent with an exponential decay. Furthermore, reliable estimation requires a minimum number of recorded spikes because the variance in estimates increases with decreasing number of non-zero activity entries~\cite{wilting_inferring_2018}.
To ensure that a given time series fulfills the requirements for reliable estimation, we implemented three consistency checks:
\begin{enumerate}
 \item Number of non-zero time bins must be at least $n_{A_t\neq 0}=1000$.
  \item Estimated timescales must be larger than a single time bin, i.e. $\hat{\tau} > \delta t$. 
 \item The exponential decay must fit the autocorrelation better than a linear function $f(\Delta t)=A\cdot\Delta t+b$, as measured by the adjusted $R^2$ of both fits.
\end{enumerate}
If a recording was not consistent with either of these requirements, it was excluded from the analysis.

\subsubsection*{Comparison with other species}
To compare the results from human MTL with other species, we re-evaluated in vivo spiking activity from macaque prefrontal cortex during a short-term memory task ($n=12$ recordings from $n=3$ macaques~\cite{pipa_performance-_2009}), and rat CA1 of the right dorsal hippocampus during a foraging task ($n=8$ recordings from $n=3$ rats \cite{rat_EC_data, mizuseki_theta_2009}). \\
The recordings from rats were obtained from the CRCNS data sharing website.
The rat experimental protocols were approved by the Institutional Animal Care and Use Committee of Rutgers University~\cite{rat_EC_data, mizuseki_theta_2009}. The macaque experiments were performed according to the German Law for the Protection of Experimental Animals, and were approved by the Regierungspräsidium Darmstadt~\cite{pipa_performance-_2009}. The procedures also conformed to the regulations issued by the NIH and the Society for Neuroscience~\cite{pipa_performance-_2009}.\\
For all spike recordings, we used a bin size $\delta t=4$~ms. The maximum time delay $\Delta t_{\text{max}}$ of the autocorrelation function was chosen separately for each species, assuring that $\Delta t_{\text{max}}$ was always in the order of multiple intrinsic network timescales ($\Delta t_{\text{max}}(\text{macaque})=2000$~ms, $\Delta t_{\text{max}}(\text{rat})=3200$~ms). We compared intrinsic timescales of humans to those of rat and macaque using the Mann-Whitney-U test. The published intrinsic timescales of the different areas in macaque~\cite{murray_hierarchy_2014} were obtained from the table provided in the paper.\\

\subsubsection*{Comparison of sleep stages}
For each patient, we analyzed one hour of sleep, where each 30-second-segment had been assigned a sleep stage. To obtain recordings of equal and sufficient length, we estimated intrinsic timescales $\hat{\tau}$ on blocks of three subsequent 30-second-segments that have been classified to the same sleep stage. This corresponds to recording segments of 90~seconds. \\
Intrinsic timescales of different sleep stages were compared using a mixed effect ANOVA with subjectID as random effect. As the distribution of $\tau$ was not consistent with a normal distribution, but rather resembled a lognormal distribution, we performed the ANOVAs on the log-transformed timescales. Specifically, we used the python interface \emph{pymer4}~\cite{jolly2018pymer4} to the R-function \emph{lmer}~\cite{bates2014fitting}, with model specification $\log{\tau} \sim  \mathrm{sleep stage}  + (1|\mathrm{subjectID})$.

\subsubsection*{Comparison of brain areas}
Intracranial recordings span hippocampus (H), amygdala (A), parahippocampal cortex (PHC) and entorhinal cortex (EC). To compare intrinsic timescales of the subregions, all spikes recorded in one subregion were combined and binned, yielding the activity in the respective subregion. Intrinsic timescales were estimated separately for each recording and subregion. Intrinsic timescales of different subregions were compared using using a mixed effect ANOVA with subjectID as random effect, using the log-transformed timescales. Specifically, we again used the python interface \emph{pymer4}~\cite{jolly2018pymer4} to the R-function \emph{lmer}~\cite{bates2014fitting}, with model specification $\log{\tau} \sim  \mathrm{subregion}  + (1|\mathrm{subjectID})$.

\clearpage

\section{Supplementary Material}

\begin{table*}[h!]
  \centering 
  \begin{tabular}{p{1.2cm}p{1.7cm}p{2.5cm}}
    \toprule
    subject \par ID &  number of\par recordings & brain regions\\
    \midrule
    1 & 4 & A, EC, PHC \\ 
    2 & 2 & H, EC \\ 
    3 & 3 & H, A, PHC \\ 
    4 & 5 & H, A, EC, PHC \\ 
    5 & 3 & H, A, PHC \\ 
    6 & 3 & H, A, EC, PHC\\ 
    7 & 2 & H, A, EC, PHC \\ 
    8 & 1 & H, A, EC\\ 
    9 & 4 & H, A, EC, PHC\\
    10 & 4 & H, A, EC, PHC \\ 
    11 & 4 & H, A\\ 
    12 & 9 & H, A, EC, PHC\\ 
    13 & 6 & H, A, EC, PHC \\ 
    14 & 4 & H, A\\ 
    15 & 7 & H, A, EC, PHC\\ 
    16 & 1 & H\\ 
    17 & 4 & H, A, EC, PHC\\ 
    18 & 9 & H, A, PHC\\ 
    19 & 4 & A, PHC \\ 
    20 & 26 & H, A, PHC\\ 
    \bottomrule
  \end{tabular}
\caption{Dataset 1: Intracranial, seizure-free recordings from subjects with medically intractable epilepsy. Recordings span different subregions of MTL, including hippocampus (H), amygdala (A), parahippocampal cortex (PHC) and entorhinal cortex (EC). When available, we evaluated recorded spikes from both hemispheres. For the comparison of subareas, spikes were binned separately for each subregion. Note, that some subregions are missing in several subjects, which indicates that too few spikes could be extracted from the respective electrodes. Patient numbers were re-assigned continuously as compared to the original dataset. Dataset provided by the Department of Epileptology in Bonn.}
\label{tab:rec_overview}
\end{table*}

\begin{table*}[h!]
  \centering 
  \begin{tabular}{p{1.2cm}p{1.7cm}p{1.2cm}p{1.2cm}p{1.2cm}p{1.2cm}p{1.2cm}}
    \toprule
    patient\par ID & evaluated\par hemisphere & wake & REM & stage N1 & stage N2 & stage N3 \\
    \midrule
    S1  & both &8 & 20 & 10 & 6 & 14 \\ 
    S2  & both &0 & 4 & 6 & 22 & 22  \\ 
    S3  & both &0 & 19 & 10 & 2 & 40  \\ 
    S4  & both &0 & 6 & 5 & 10 & 44  \\ 
    S5  & right  &0 & 2 & 2 & 7 & 25  \\ 
    S6  & both &0 & 4 & 5 & 16 & 14  \\ 
    \bottomrule
  \end{tabular}
  \caption{Dataset 2: Intracranial sleep recordings from epilepsy patients. Recordings span different subregions of MTL, including hippocampus (H), amygdala (A), parahippocampal cortex (PHC) and entorhinal cortex (EC). Patient numbers were re-assigned continuously as compared to the original dataset. Two patients are also in dataset 1 (table~\ref{tab:rec_overview}): patient S2 is equal to patient 15; patient S5 is equal to patient 20. When available, we evaluated recorded spikes from both hemispheres. Numbers refer to the number of 90~s recording segments in the respective sleep stage. Dataset provided by the Department of Epileptology in Bonn.}
  \label{table_sleep}
\end{table*}

\begin{figure*}
    \centering
    \includegraphics[width=0.5\textwidth]{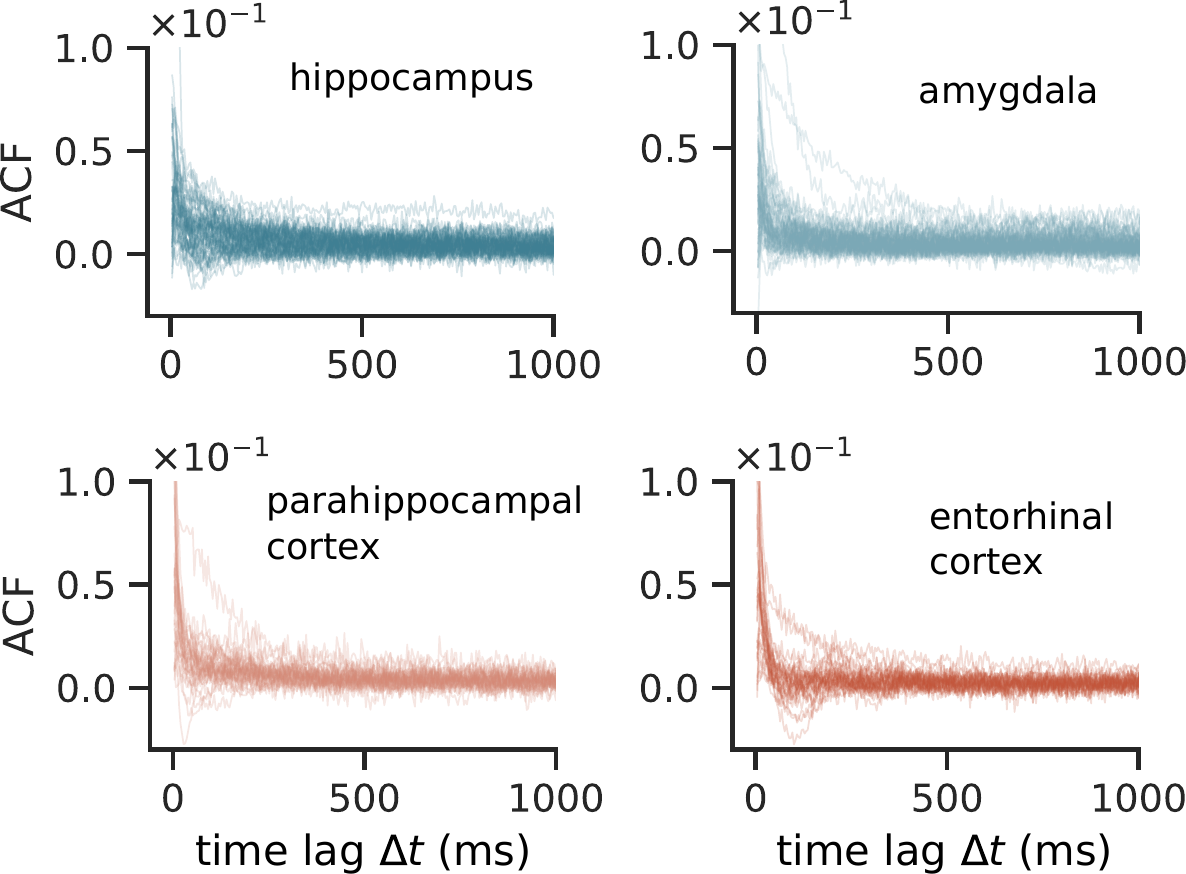}
    \caption{Overview of all autocorrelation functions (ACF) of spiking activity in different subregions of human MTL. ACFs are widely consistent with exponential decays. Even within the individual subregions, the rate of decay, which defines the intrinsic timescale $\tau$ is highly diverse.}
    \label{suppfig:autocorrelations_regions}
\end{figure*}

\begin{figure*}
    \centering
    \includegraphics[width=0.9\textwidth]{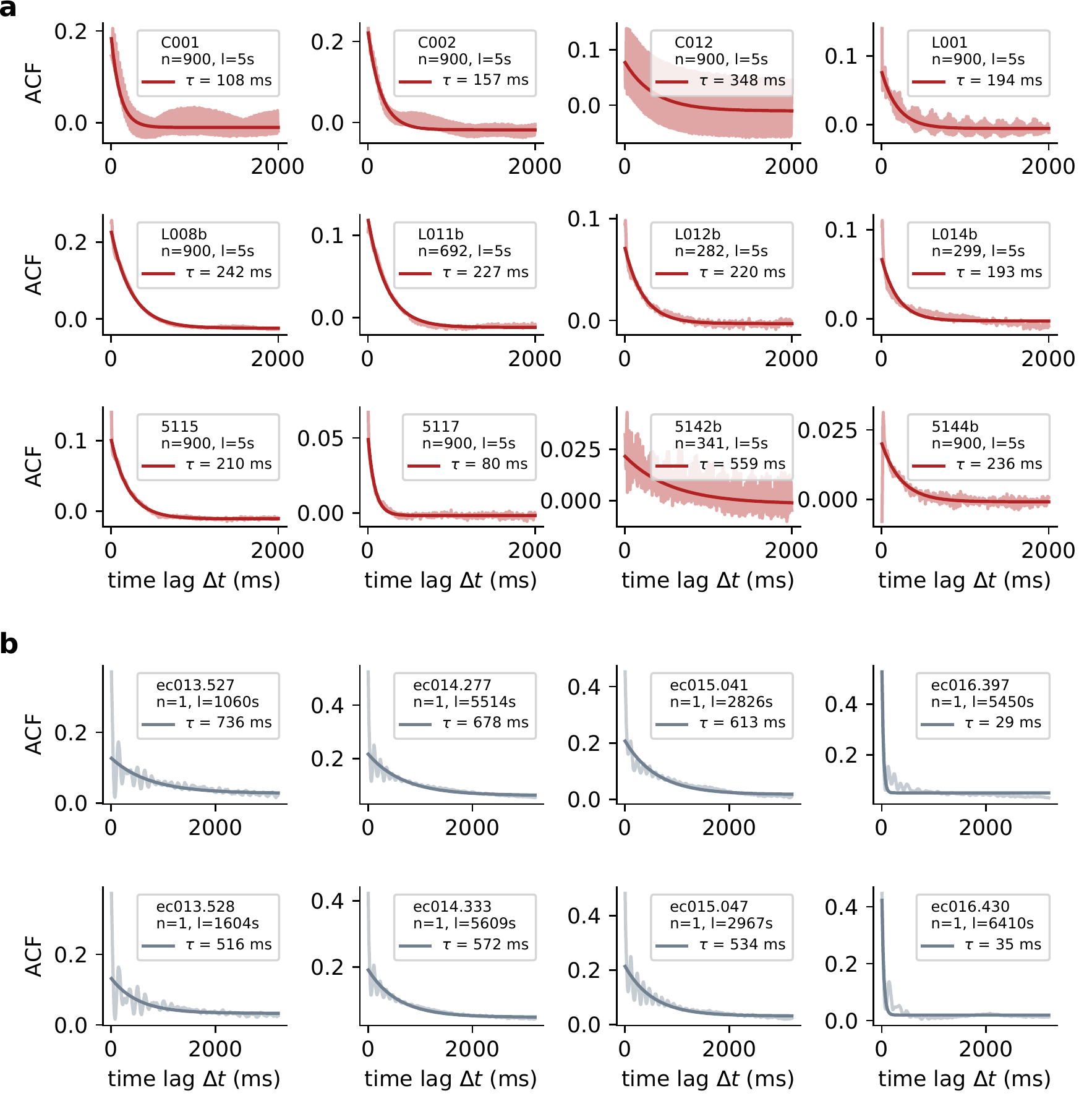}
    \caption{Autocorrelation functions (ACF) and fits of spiking activity in different species. \textbf{a} Recordings from macaque prefrontal cortex~\cite{pipa_performance-_2009}. \textbf{b} Recordings from rat hippocampus~\cite{rat_EC_data, mizuseki_theta_2009}. Note that rat EC shows clear theta-oscillations in addition to the exponential decay.}
    \label{suppfig:autocorrelations_species}
\end{figure*}

\newpage

\begin{figure*}
    \includegraphics[width=0.99\textwidth]{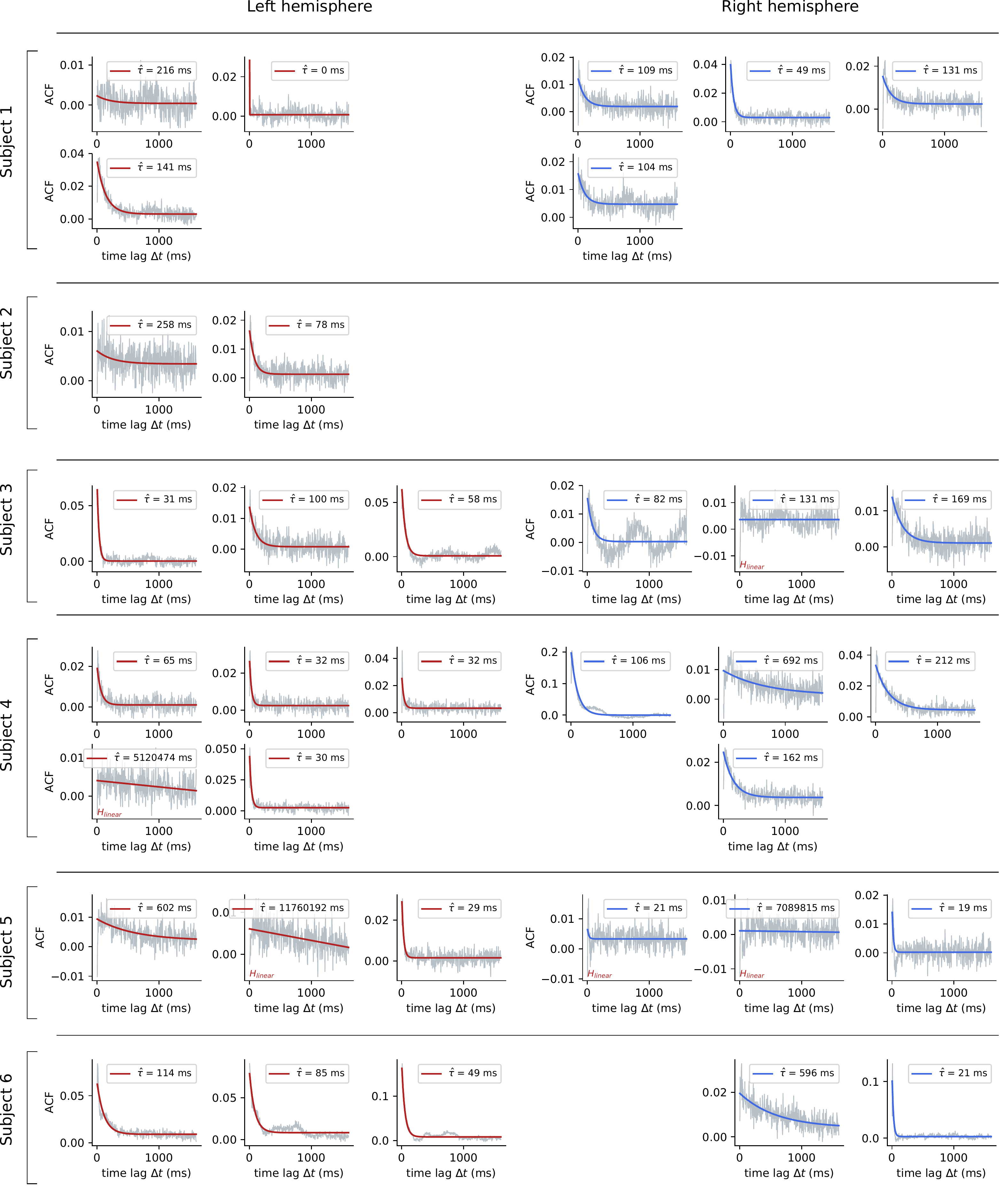}
    \caption{Autocorrelation functions (ACF) and fits of spiking activity in human MTL, subjects 1-6. When available, both hemispheres are shown, where plots in the same position belong to simultaneous recordings. Missing values plots indicate that too few spikes were registered, and red annotations show that a recording was excluded from further analysis based on one of the exclusion criteria.}
    \label{suppfig:ACFs_first}
\end{figure*}

\begin{figure*}
    \includegraphics[width=0.99\textwidth]{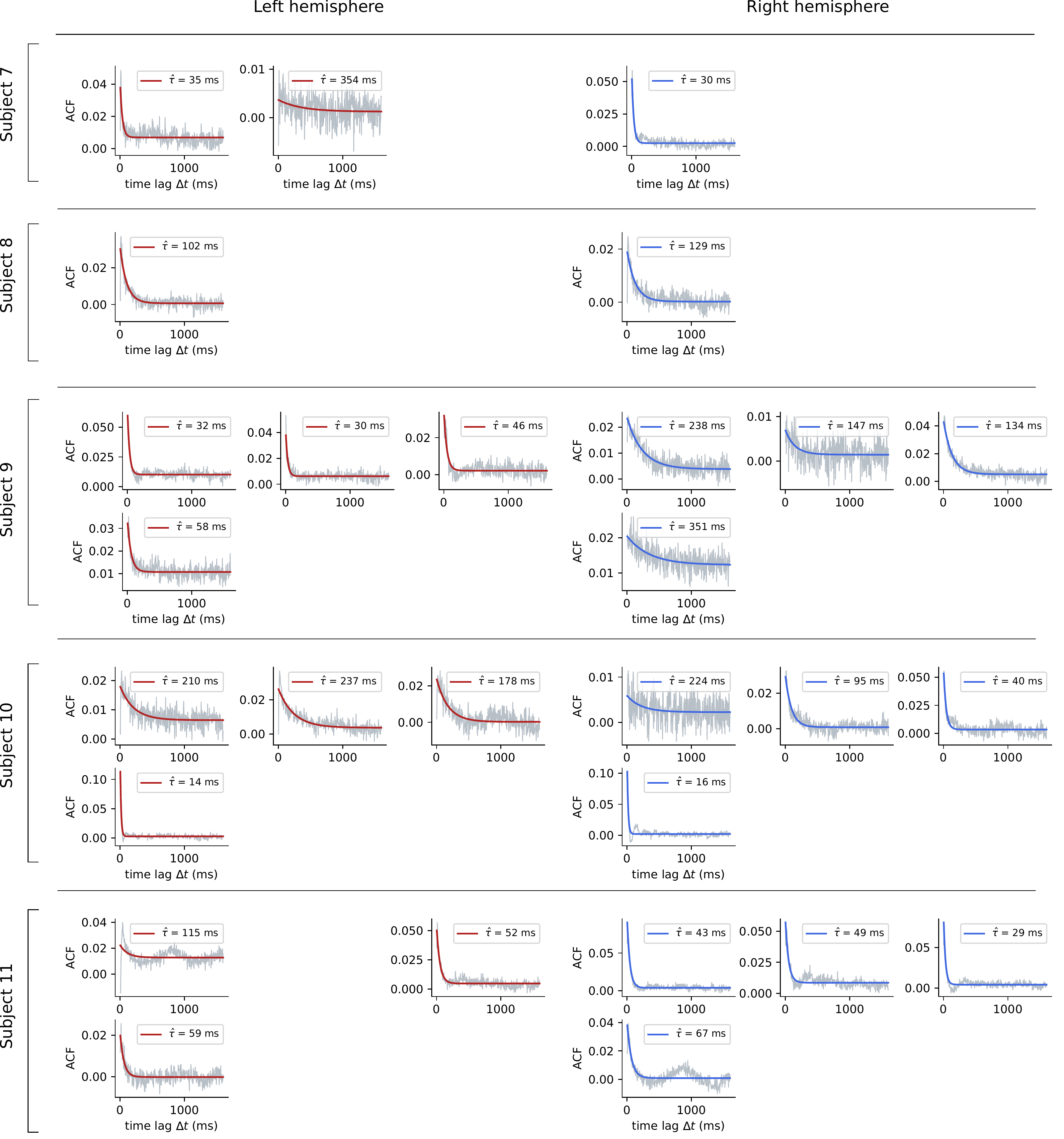}
        \caption{Autocorrelation functions (ACF) and fits of spiking activity in human MTL, subjects 7-11. When available, both hemispheres are shown, where plots in the same position belong to simultaneous recordings. Missing values plots indicate that too few spikes were registered, and red annotations show that a recording was excluded from further analysis based on one of the exclusion criteria.}
\end{figure*}

\begin{figure*}
    \includegraphics[width=0.99\textwidth]{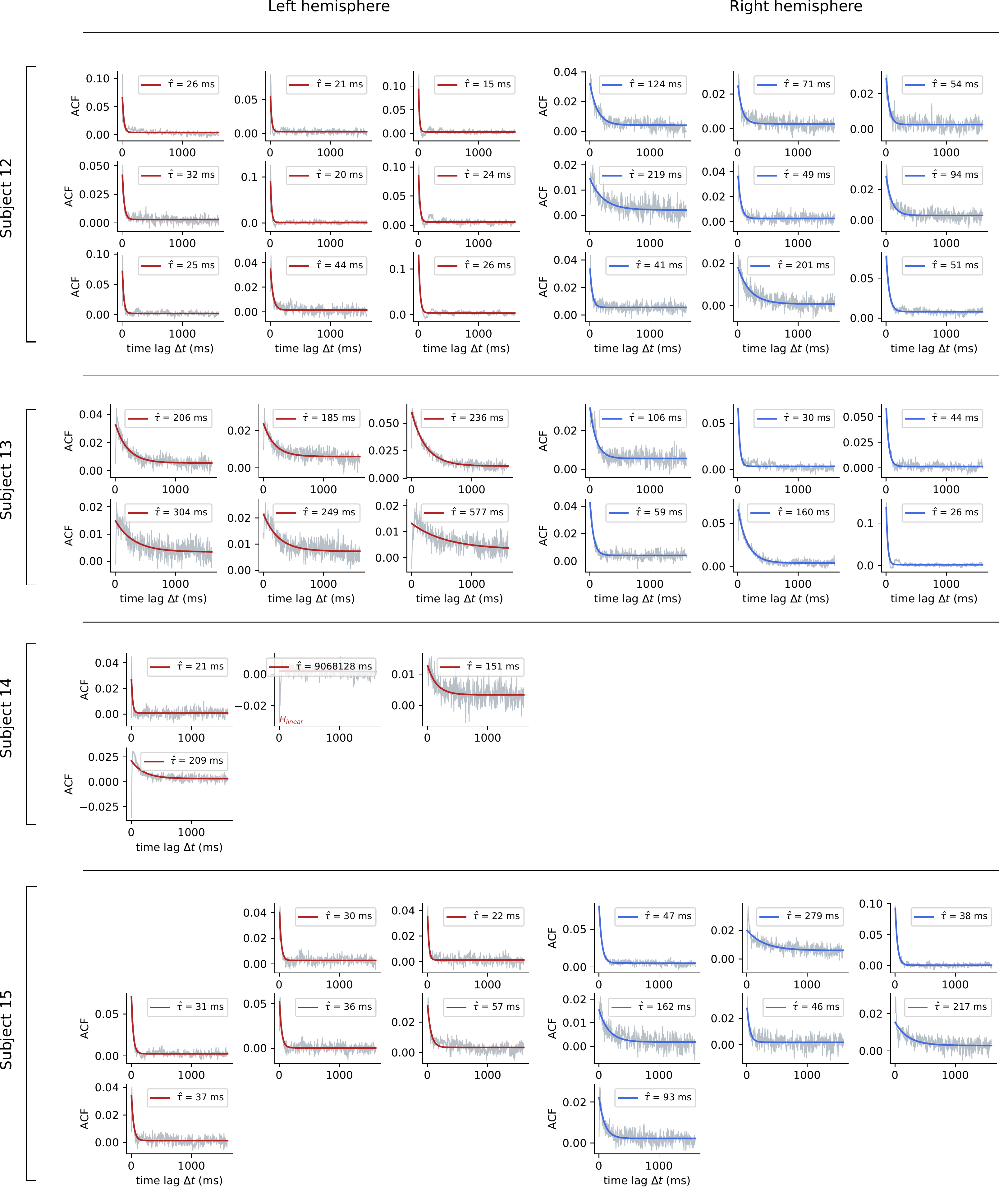}
        \caption{Autocorrelation functions (ACF) and fits of spiking activity in human MTL, subjects 12-15. When available, both hemispheres are shown, where plots in the same position belong to simultaneous recordings. Missing values plots indicate that too few spikes were registered, and red annotations show that a recording was excluded from further analysis based on one of the exclusion criteria.}
\end{figure*}

\begin{figure*}
    \includegraphics[width=0.99\textwidth]{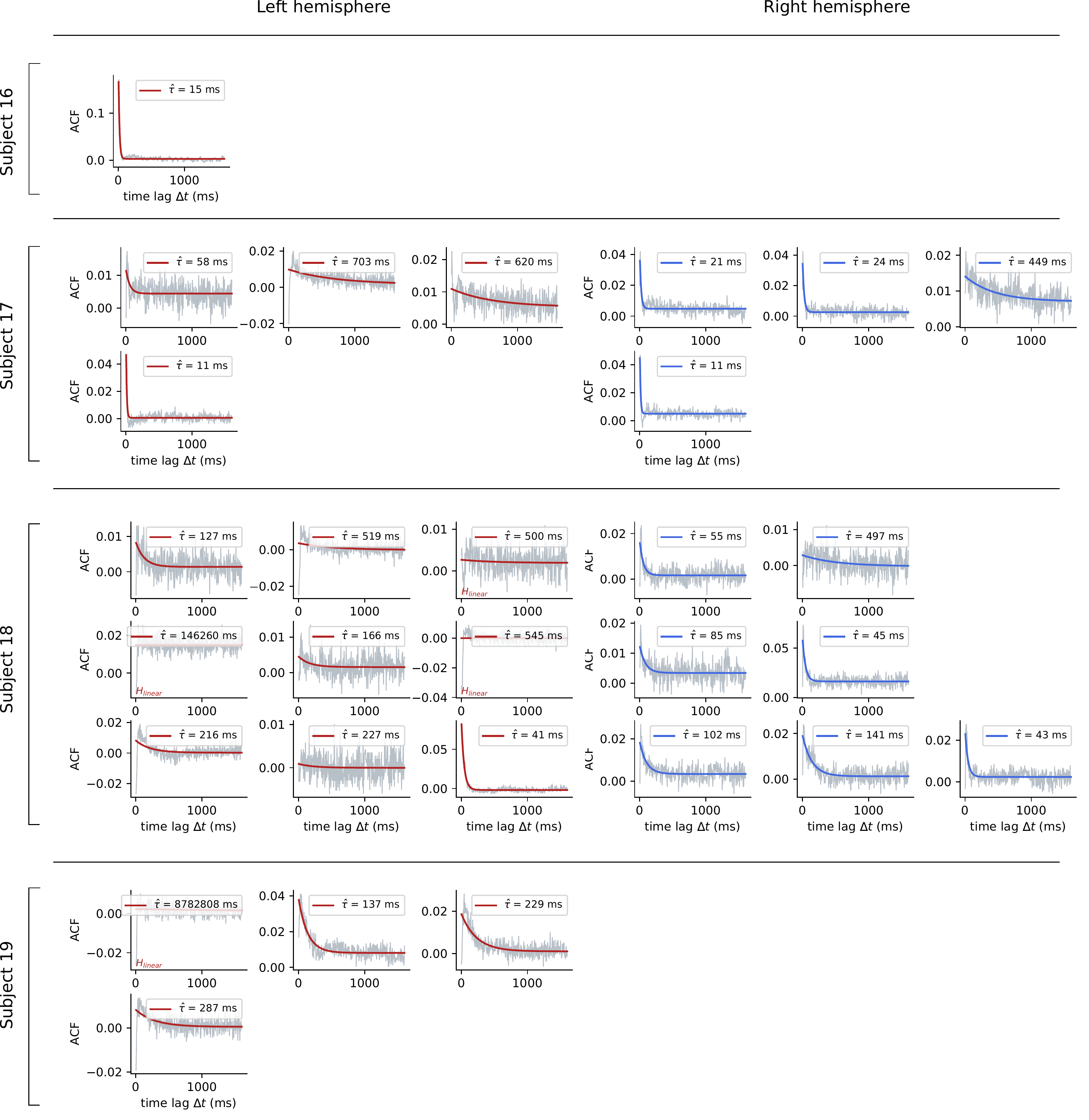}
        \caption{Autocorrelation functions (ACF) and fits of spiking activity in human MTL, subjects 16-19. When available, both hemispheres are shown, where plots in the same position belong to simultaneous recordings. Missing values plots indicate that too few spikes were registered, and red annotations show that a recording was excluded from further analysis based on one of the exclusion criteria.}
\end{figure*}

\begin{figure*}
    \includegraphics[width=0.99\textwidth]{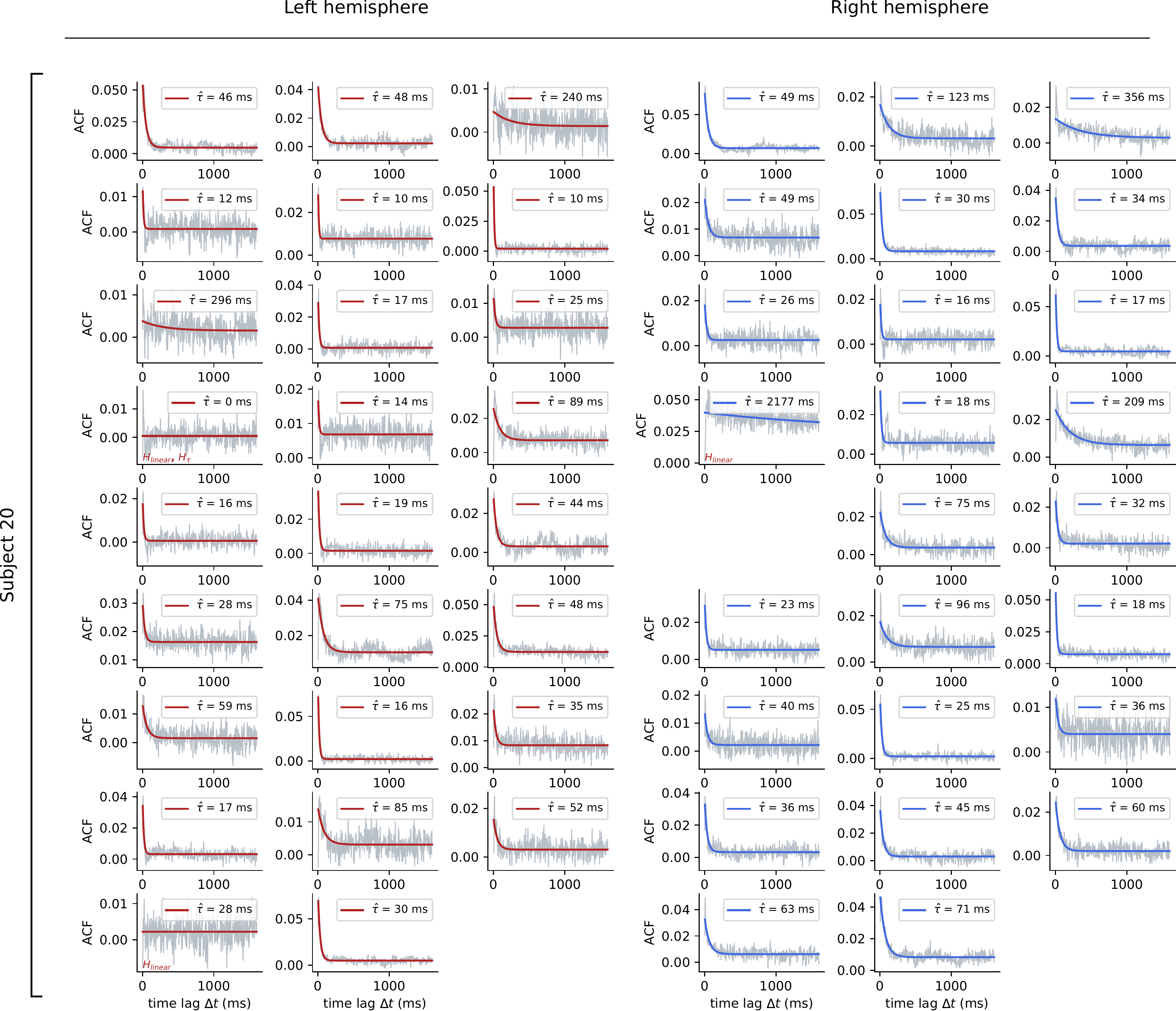}
        \caption{Autocorrelation functions (ACF) and fits of spiking activity in human MTL, subjects 20. When available, both hemispheres are shown, where plots in the same position belong to simultaneous recordings. Missing values plots indicate that too few spikes were registered, and red annotations show that a recording was excluded from further analysis based on one of the exclusion criteria.}
        \label{suppfig:ACFs_last}
\end{figure*}



\clearpage
\begin{thebibliography}{42}
\providecommand{\natexlab}[1]{#1}
\providecommand{\urlprefix}{}
\providecommand{\doiprefix}{doi: }

\bibitem[{Bates et~al.(2014)Bates, Douglas and M{\"a}chler, Martin and Bolker,
  Ben and Walker, Steve}]{bates2014fitting}
\textbf{\color{eLifeMediumGrey} Bates D}, M{\"a}chler M, Bolker B, Walker S.
\newblock Fitting linear mixed-effects models using lme4.
\newblock arXiv preprint arXiv:14065823.  2014; .

\bibitem[{Cavanagh et~al.(2020)Cavanagh, Sean E. and Hunt, Laurence T. and
  Kennerley, Steven W.}]{cavanagh_diversity_2020}
\textbf{\color{eLifeMediumGrey} Cavanagh SE}, Hunt LT, Kennerley SW.
\newblock A {Diversity} of {Intrinsic} {Timescales} {Underlie} {Neural}
  {Computations}.
\newblock Frontiers in Neural Circuits.  2020; 14.
\newblock
  \urlprefix\url{https://www.frontiersin.org/article/10.3389/fncir.2020.615626}.

\bibitem[{Chaudhuri et~al.(2015)Chaudhuri, Rishidev and Knoblauch, Kenneth and
  Gariel, Marie-Alice and Kennedy, Henry and Wang,
  Xiao-Jing}]{chaudhuri_large-scale_2015}
\textbf{\color{eLifeMediumGrey} Chaudhuri R}, Knoblauch K, Gariel MA, Kennedy
  H, Wang XJ.
\newblock A {Large}-{Scale} {Circuit} {Mechanism} for {Hierarchical}
  {Dynamical} {Processing} in the {Primate} {Cortex}.
\newblock Neuron.  2015 Oct; 88(2):419--431.
\newblock
  \urlprefix\url{http://www.sciencedirect.com/science/article/pii/S0896627315007655},
  \href{10.1016/j.neuron.2015.09.008}{\doiprefix
  \detokenize{10.1016/j.neuron.2015.09.008}}.

\bibitem[{Cirillo et~al.(2018)Cirillo, Rossella and Fascianelli, Valeria and
  Ferrucci, Lorenzo and Genovesio, Aldo}]{cirillo2018neural}
\textbf{\color{eLifeMediumGrey} Cirillo R}, Fascianelli V, Ferrucci L,
  Genovesio A.
\newblock Neural intrinsic timescales in the macaque dorsal premotor cortex
  predict the strength of spatial response coding.
\newblock iScience.  2018; 10:203--210.

\bibitem[{Cramer et~al.(2020)Cramer, Benjamin and St{\"o}ckel, David and Kreft,
  Markus and Wibral, Michael and Schemmel, Johannes and Meier, Karlheinz and
  Priesemann, Viola}]{cramer2020control}
\textbf{\color{eLifeMediumGrey} Cramer B}, St{\"o}ckel D, Kreft M, Wibral M,
  Schemmel J, Meier K, Priesemann V.
\newblock Control of criticality and computation in spiking neuromorphic
  networks with plasticity.
\newblock Nature communications.  2020; 11(1):1--11.

\bibitem[{Gao et~al.(2020)Gao, Richard and van den Brink, Ruud L and Pfeffer,
  Thomas and Voytek, Bradley}]{gao2020neuronal}
\textbf{\color{eLifeMediumGrey} Gao R}, van~den Brink RL, Pfeffer T, Voytek B.
\newblock Neuronal timescales are functionally dynamic and shaped by cortical
  microarchitecture.
\newblock Elife.  2020; 9:e61277.

\bibitem[{Golesorkhi et~al.(2021)Golesorkhi, Mehrshad and Gomez-Pilar, Javier
  and Zilio, Federico and Berberian, Nareg and Wolff, Annemarie and Yagoub,
  Mustapha C. E. and Northoff, Georg}]{golesorkhi_brain_2021}
\textbf{\color{eLifeMediumGrey} Golesorkhi M}, Gomez-Pilar J, Zilio F,
  Berberian N, Wolff A, Yagoub MCE, Northoff G.
\newblock The brain and its time: intrinsic neural timescales are key for input
  processing.
\newblock Communications Biology.  2021 Aug; 4(1):1--16.
\newblock \urlprefix\url{https://www.nature.com/articles/s42003-021-02483-6},
  \href{10.1038/s42003-021-02483-6}{\doiprefix
  \detokenize{10.1038/s42003-021-02483-6}}.

\bibitem[{Hagemann et~al.(2021)Hagemann, Annika and Wilting, Jens and
  Samimizad, Bita and Mormann, Florian and Priesemann,
  Viola}]{hagemann2021assessing}
\textbf{\color{eLifeMediumGrey} Hagemann A}, Wilting J, Samimizad B, Mormann F,
  Priesemann V.
\newblock Assessing criticality in pre-seizure single-neuron activity of human
  epileptic cortex.
\newblock PLoS computational biology.  2021; 17(3):e1008773.

\bibitem[{Hasson et~al.(2015)Hasson, Uri and Chen, Janice and Honey,
  Christopher J}]{hasson2015hierarchical}
\textbf{\color{eLifeMediumGrey} Hasson U}, Chen J, Honey CJ.
\newblock Hierarchical process memory: memory as an integral component of
  information processing.
\newblock Trends in cognitive sciences.  2015; 19(6):304--313.

\bibitem[{Hasson et~al.(2008)Hasson, Uri and Yang, Eunice and Vallines, Ignacio
  and Heeger, David J and Rubin, Nava}]{hasson2008hierarchy}
\textbf{\color{eLifeMediumGrey} Hasson U}, Yang E, Vallines I, Heeger DJ, Rubin
  N.
\newblock A hierarchy of temporal receptive windows in human cortex.
\newblock Journal of Neuroscience.  2008; 28(10):2539--2550.

\bibitem[{Honey et~al.(2012)Honey, Christopher J and Thesen, Thomas and Donner,
  Tobias H and Silbert, Lauren J and Carlson, Chad E and Devinsky, Orrin and
  Doyle, Werner K and Rubin, Nava and Heeger, David J and Hasson,
  Uri}]{honey2012slow}
\textbf{\color{eLifeMediumGrey} Honey CJ}, Thesen T, Donner TH, Silbert LJ,
  Carlson CE, Devinsky O, Doyle WK, Rubin N, Heeger DJ, Hasson U.
\newblock Slow cortical dynamics and the accumulation of information over long
  timescales.
\newblock Neuron.  2012; 76(2):423--434.

\bibitem[{Huang and Doiron(2017)Huang, Chengcheng and Doiron,
  Brent}]{huang_once_2017}
\textbf{\color{eLifeMediumGrey} Huang C}, Doiron B.
\newblock Once upon a (slow) time in the land of recurrent neuronal
  networks….
\newblock Current Opinion in Neurobiology.  2017 Oct; 46:31--38.
\newblock
  \urlprefix\url{http://www.sciencedirect.com/science/article/pii/S0959438817300193},
  \href{10.1016/j.conb.2017.07.003}{\doiprefix
  \detokenize{10.1016/j.conb.2017.07.003}}.

\bibitem[{Iber(2007)Iber, Conrad}]{iber2007aasm}
\textbf{\color{eLifeMediumGrey} Iber C}.
\newblock The AASM manual for the scoring of sleep and associated events:
  Rules.
\newblock Terminology and Technical Specification.  2007; .

\bibitem[{Jaeger(2001)Jaeger, Herbert}]{jaeger2001echo}
\textbf{\color{eLifeMediumGrey} Jaeger H}.
\newblock The echo state approach to analysing and training recurrent neural
  networks-with an erratum note.
\newblock Bonn, Germany: German National Research Center for Information
  Technology GMD Technical Report.  2001; 148(34):13.

\bibitem[{Jolly(2018)Jolly, Eshin}]{jolly2018pymer4}
\textbf{\color{eLifeMediumGrey} Jolly E}.
\newblock Pymer4: connecting R and Python for linear mixed modeling.
\newblock Journal of Open Source Software.  2018; 3(31):862.

\bibitem[{Levina and Priesemann(2017)Levina, A. and Priesemann,
  V.}]{levina_subsampling_2017}
\textbf{\color{eLifeMediumGrey} Levina A}, Priesemann V.
\newblock Subsampling scaling.
\newblock Nature Communications.  2017 May; 8:15140.
\newblock \urlprefix\url{http://www.nature.com/doifinder/10.1038/ncomms15140},
  \href{10.1038/ncomms15140}{\doiprefix \detokenize{10.1038/ncomms15140}}.

\bibitem[{Loidolt et~al.(2020)Loidolt, Matthias and Rudelt, Lucas and
  Priesemann, Viola}]{loidolt_sequence_2020}
\textbf{\color{eLifeMediumGrey} Loidolt M}, Rudelt L, Priesemann V.
\newblock Sequence Memory in Recurrent Neuronal Network Can Develop without
  Structured Input.
\newblock bioRxiv.  2020 Sep; p. 2020.09.15.297580.
\newblock \href{10.1101/2020.09.15.297580}{\doiprefix
  \detokenize{10.1101/2020.09.15.297580}}.

\bibitem[{Maass et~al.(2002)Maass, Wolfgang and Natschl{\"a}ger, Thomas and
  Markram, Henry}]{maass2002real}
\textbf{\color{eLifeMediumGrey} Maass W}, Natschl{\"a}ger T, Markram H.
\newblock Real-time computing without stable states: A new framework for neural
  computation based on perturbations.
\newblock Neural computation.  2002; 14(11):2531--2560.

\bibitem[{Meisel(2019)Meisel, Christian}]{meisel_antiepileptic_2019}
\textbf{\color{eLifeMediumGrey} Meisel C}.
\newblock Antiepileptic drugs induce subcritical dynamics in human cortical
  networks.
\newblock arXiv:190413026 [q-bio].  2019 Apr;
  \urlprefix\url{http://arxiv.org/abs/1904.13026}, arXiv: 1904.13026.

\bibitem[{Meisel et~al.(2017)Meisel, Christian and Klaus, Andreas and
  Vyazovskiy, Vladyslav V and Plenz, Dietmar}]{meisel2017interplay}
\textbf{\color{eLifeMediumGrey} Meisel C}, Klaus A, Vyazovskiy VV, Plenz D.
\newblock The interplay between long-and short-range temporal correlations
  shapes cortex dynamics across vigilance states.
\newblock Journal of neuroscience.  2017; 37(42):10114--10124.

\bibitem[{Meisel et~al.(2015)Meisel, Christian and Schulze-Bonhage, Andreas and
  Freestone, Dean and Cook, Mark James and Achermann, Peter and Plenz,
  Dietmar}]{meisel2015intrinsic}
\textbf{\color{eLifeMediumGrey} Meisel C}, Schulze-Bonhage A, Freestone D, Cook
  MJ, Achermann P, Plenz D.
\newblock Intrinsic excitability measures track antiepileptic drug action and
  uncover increasing/decreasing excitability over the wake/sleep cycle.
\newblock Proceedings of the National Academy of Sciences.  2015;
  112(47):14694--14699.

\bibitem[{Mizuseki et~al.(2009)Mizuseki, Kenji and Sirota, Anton and
  Pastalkova, Eva and Buzs\'{a}ki, Gy\"orgy}]{mizuseki_theta_2009}
\textbf{\color{eLifeMediumGrey} Mizuseki K}, Sirota A, Pastalkova E,
  Buzs\'{a}ki G.
\newblock Theta oscillations provide temporal windows for local circuit
  computation in the entorhinal-hippocampal loop.
\newblock Neuron.  2009 Oct; 64(2):267--280.
\newblock \href{10.1016/j.neuron.2009.08.037}{\doiprefix
  \detokenize{10.1016/j.neuron.2009.08.037}}.

\bibitem[{Mizuseki~K(2009)Mizuseki K, Sirota A, Pastalkova E, Buzs\'{a}ki
  G.}]{rat_EC_data}
\textbf{\color{eLifeMediumGrey} Mizuseki~K PEBG Sirota~A}, Multi-unit
  recordings from the rat hippocampus made during open field foraging.; 2009.
\newblock \urlprefix\url{http://dx.doi.org/10.6080/K0Z60KZ9}, available at
  \url{http://dx.doi.org/10.6080/K0Z60KZ9}. Accessed 07.11.2019.

\bibitem[{Murray et~al.(2014)Murray, John D. and Bernacchia, Alberto and
  Freedman, David J. and Romo, Ranulfo and Wallis, Jonathan D. and Cai, Xinying
  and Padoa-Schioppa, Camillo and Pasternak, Tatiana and Seo, Hyojung and Lee,
  Daeyeol and Wang, Xiao-Jing}]{murray_hierarchy_2014}
\textbf{\color{eLifeMediumGrey} Murray JD}, Bernacchia A, Freedman DJ, Romo R,
  Wallis JD, Cai X, Padoa-Schioppa C, Pasternak T, Seo H, Lee D, Wang XJ.
\newblock A hierarchy of intrinsic timescales across primate cortex.
\newblock Nature Neuroscience.  2014 Dec; 17(12):1661--1663.
\newblock \urlprefix\url{https://www.nature.com/articles/nn.3862},
  \href{10.1038/nn.3862}{\doiprefix \detokenize{10.1038/nn.3862}}.

\bibitem[{Neto et~al.(2020)Neto, Joao Pinheiro and Spitzner, Franz Paul and
  Priesemann, Viola}]{neto_unified_2020}
\textbf{\color{eLifeMediumGrey} Neto JP}, Spitzner FP, Priesemann V.
\newblock A unified picture of neuronal avalanches arises from the
  understanding of sampling effects.
\newblock arXiv:191009984 [cond-mat, physics:nlin, physics:physics, q-bio].
  2020 Mar; \urlprefix\url{http://arxiv.org/abs/1910.09984}, arXiv: 1910.09984.

\bibitem[{Niediek et~al.(2016)Niediek, Johannes and Bostr\"om, Jan and Elger,
  Christian E. and Mormann, Florian}]{niediek_reliable_2016}
\textbf{\color{eLifeMediumGrey} Niediek J}, Bostr\"om J, Elger CE, Mormann F.
\newblock Reliable {Analysis} of {Single}-{Unit} {Recordings} from the {Human}
  {Brain} under {Noisy} {Conditions}: {Tracking} {Neurons} over {Hours}.
\newblock PLoS ONE.  2016 Dec; 11(12).
\newblock
  \urlprefix\url{https://www.ncbi.nlm.nih.gov/pmc/articles/PMC5145161/},
  \href{10.1371/journal.pone.0166598}{\doiprefix
  \detokenize{10.1371/journal.pone.0166598}}.

\bibitem[{Pipa et~al.(2009)Pipa, Gordon and Staedtler, Ellen S. and Rodriguez,
  Eugenio F. and Waltz, James A. and Muckli, Lars and Singer, Wolf and Goebel,
  Rainer and Munk, Matthias H. J.}]{pipa_performance-_2009}
\textbf{\color{eLifeMediumGrey} Pipa G}, Staedtler ES, Rodriguez EF, Waltz JA,
  Muckli L, Singer W, Goebel R, Munk MHJ.
\newblock Performance- and stimulus-dependent oscillations in monkey prefrontal
  cortex during short-term memory.
\newblock Frontiers in Integrative Neuroscience.  2009; 3.
\newblock
  \urlprefix\url{https://www.frontiersin.org/articles/10.3389/neuro.07.025.2009/full},
  \href{10.3389/neuro.07.025.2009}{\doiprefix
  \detokenize{10.3389/neuro.07.025.2009}}.

\bibitem[{Priesemann et~al.(2009)Priesemann, Viola and Munk, Matthias HJ and
  Wibral, Michael}]{priesemann2009subsampling}
\textbf{\color{eLifeMediumGrey} Priesemann V}, Munk MH, Wibral M.
\newblock Subsampling effects in neuronal avalanche distributions recorded in
  vivo.
\newblock BMC neuroscience.  2009; 10(1):40.

\bibitem[{Priesemann et~al.(2013)Priesemann, Viola and Valderrama, Mario and
  Wibral, Michael and Quyen, Michel Le Van}]{priesemann_neuronal_2013}
\textbf{\color{eLifeMediumGrey} Priesemann V}, Valderrama M, Wibral M, Quyen
  MLV.
\newblock Neuronal {Avalanches} {Differ} from {Wakefulness} to {Deep} {Sleep}
  – {Evidence} from {Intracranial} {Depth} {Recordings} in {Humans}.
\newblock PLOS Computational Biology.  2013 Mar; 9(3):e1002985.
\newblock
  \urlprefix\url{http://journals.plos.org/ploscompbiol/article?id=10.1371/journal.pcbi.1002985},
  \href{10.1371/journal.pcbi.1002985}{\doiprefix
  \detokenize{10.1371/journal.pcbi.1002985}}.

\bibitem[{Priesemann et~al.(2014)Priesemann, Viola and Wibral, Michael and
  Valderrama, Mario and Pr\"opper, Robert and Le Van Quyen, Michel and Geisel,
  Theo and Triesch, Jochen and Nikolić, Danko and Munk, Matthias H.
  J.}]{priesemann_spike_2014}
\textbf{\color{eLifeMediumGrey} Priesemann V}, Wibral M, Valderrama M,
  Pr\"opper R, Le~Van~Quyen M, Geisel T, Triesch J, Nikolić D, Munk MHJ.
\newblock Spike avalanches in vivo suggest a driven, slightly subcritical brain
  state.
\newblock Frontiers in Systems Neuroscience.  2014; 8.
\newblock
  \urlprefix\url{https://www.frontiersin.org/articles/10.3389/fnsys.2014.00108/full},
  \href{10.3389/fnsys.2014.00108}{\doiprefix
  \detokenize{10.3389/fnsys.2014.00108}}.

\bibitem[{Rasch and Born(2013)Rasch, Björn and Born, Jan}]{rasch_about_2013}
\textbf{\color{eLifeMediumGrey} Rasch B}, Born J.
\newblock About {Sleep}'s {Role} in {Memory}.
\newblock Physiological Reviews.  2013 Apr; 93(2):681--766.
\newblock
  \urlprefix\url{https://journals.physiology.org/doi/full/10.1152/physrev.00032.2012},
  \href{10.1152/physrev.00032.2012}{\doiprefix
  \detokenize{10.1152/physrev.00032.2012}}.

\bibitem[{Rudelt et~al.(2022)Rudelt, Lucas and Gonzalez Marx, Daniel and
  Cramer, Benjamin and Priesemann, Viola}]{rudelt2022mouse}
\textbf{\color{eLifeMediumGrey} Rudelt L}, Gonzalez~Marx D, Cramer B,
  Priesemann V.
\newblock Signatures of hierarchical temporal processing in the mouse visual
  system; 2022, in preparation.

\bibitem[{Shafiei et~al.(2020)Shafiei, Golia and Markello, Ross D and Vos de
  Wael, Reinder and Bernhardt, Boris C and Fulcher, Ben D and Misic,
  Bratislav}]{shafiei_topographic_2020}
\textbf{\color{eLifeMediumGrey} Shafiei G}, Markello RD, Vos~de Wael R,
  Bernhardt BC, Fulcher BD, Misic B.
\newblock Topographic gradients of intrinsic dynamics across neocortex.
\newblock eLife.  2020 Dec; 9:e62116.
\newblock \urlprefix\url{https://doi.org/10.7554/eLife.62116},
  \href{10.7554/eLife.62116}{\doiprefix \detokenize{10.7554/eLife.62116}}.

\bibitem[{Siegle et~al.(2019)Siegle, Joshua H. and Jia, Xiaoxuan and Durand,
  S\'{e}verine and Gale, Sam and Bennett, Corbett and Graddis, Nile and Heller,
  Greggory and Ramirez, Tamina K. and Choi, Hannah and Luviano, Jennifer A. and
  Groblewski, Peter A. and Ahmed, Ruweida and Arkhipov, Anton and Bernard, Amy
  and Billeh, Yazan N. and Brown, Dillan and Buice, Michael A. and Cain,
  Nicolas and Caldejon, Shiella and Casal, Linzy and Cho, Andrew and Chvilicek,
  Maggie and Cox, Timothy C. and Dai, Kael and Denman, Daniel J. and de Vries,
  Saskia E. J. and Dietzman, Roald and Esposito, Luke and Farrell, Colin and
  Feng, David and Galbraith, John and Garrett, Marina and Gelfand, Emily C. and
  Hancock, Nicole and Harris, Julie A. and Howard, Robert and Hu, Brian and
  Hytnen, Ross and Iyer, Ramakrishnan and Jessett, Erika and Johnson, Katelyn
  and Kato, India and Kiggins, Justin and Lambert, Sophie and Lecoq, Jerome and
  Ledochowitsch, Peter and Lee, Jung Hoon and Leon, Arielle and Li, Yang and
  Liang, Elizabeth and Long, Fuhui and Mace, Kyla and Melchior, Jose and
  Millman, Daniel and Mollenkopf, Tyler and Nayan, Chelsea and Ng, Lydia and
  Ngo, Kiet and Nguyen, Thuyahn and Nicovich, Philip R. and North, Kat and
  Ocker, Gabriel Koch and Ollerenshaw, Doug and Oliver, Michael and Pachitariu,
  Marius and Perkins, Jed and Reding, Melissa and Reid, David and Robertson,
  Miranda and Ronellenfitch, Kara and Seid, Sam and Slaughterbeck, Cliff and
  Stoecklin, Michelle and Sullivan, David and Sutton, Ben and Swapp, Jackie and
  Thompson, Carol and Turner, Kristen and Wakeman, Wayne and Whitesell,
  Jennifer D. and Williams, Derric and Williford, Ali and Young, Rob and Zeng,
  Hongkui and Naylor, Sarah and Phillips, John W. and Reid, R. Clay and
  Mihalas, Stefan and Olsen, Shawn R. and Koch, Christof}]{siegle_survey_2019}
\textbf{\color{eLifeMediumGrey} Siegle JH}, Jia X, Durand S, Gale S, Bennett C,
  Graddis N, Heller G, Ramirez TK, Choi H, Luviano JA, Groblewski PA, Ahmed R,
  Arkhipov A, Bernard A, Billeh YN, Brown D, Buice MA, Cain N, Caldejon S,
  Casal L, et~al.
\newblock A survey of spiking activity reveals a functional hierarchy of mouse
  corticothalamic visual areas.
\newblock .  2019 Oct;
  \urlprefix\url{http://biorxiv.org/lookup/doi/10.1101/805010},
  \href{10.1101/805010}{\doiprefix \detokenize{10.1101/805010}}.

\bibitem[{Spitzner et~al.(2021)Spitzner, F Paul and Dehning, Jonas and Wilting,
  Jens and Hagemann, Annika and P. Neto, J and Zierenberg, Johannes and
  Priesemann, Viola}]{spitzner2021mr}
\textbf{\color{eLifeMediumGrey} Spitzner FP}, Dehning J, Wilting J, Hagemann A,
  P~Neto J, Zierenberg J, Priesemann V.
\newblock MR. Estimator, a toolbox to determine intrinsic timescales from
  subsampled spiking activity.
\newblock Plos one.  2021; 16(4):e0249447.

\bibitem[{Wasmuht et~al.(2018)Wasmuht, D. F. and Spaak, E. and Buschman, T. J.
  and Miller, E. K. and Stokes, M. G.}]{wasmuht_intrinsic_2018}
\textbf{\color{eLifeMediumGrey} Wasmuht DF}, Spaak E, Buschman TJ, Miller EK,
  Stokes MG.
\newblock Intrinsic neuronal dynamics predict distinct functional roles during
  working memory.
\newblock Nature Communications.  2018 Dec; 9(1):3499.
\newblock \urlprefix\url{http://www.nature.com/articles/s41467-018-05961-4},
  \href{10.1038/s41467-018-05961-4}{\doiprefix
  \detokenize{10.1038/s41467-018-05961-4}}.

\bibitem[{Watanabe et~al.(2019)Watanabe, Takamitsu and Rees, Geraint and
  Masuda, Naoki}]{watanabe_atypical_2019}
\textbf{\color{eLifeMediumGrey} Watanabe T}, Rees G, Masuda N.
\newblock Atypical intrinsic neural timescale in autism.
\newblock eLife.  2019 Feb; 8:e42256.
\newblock \urlprefix\url{https://doi.org/10.7554/eLife.42256},
  \href{10.7554/eLife.42256}{\doiprefix \detokenize{10.7554/eLife.42256}}.

\bibitem[{Wilting and Priesemann(2019{\natexlab{a}})Wilting, J and Priesemann,
  V}]{Wilting2019_25}
\textbf{\color{eLifeMediumGrey} Wilting J}, Priesemann V.
\newblock 25 years of criticality in neuroscience — established results, open
  controversies, novel concepts.
\newblock Current Opinion in Neurobiology.  2019 Oct; 58:105--111.
\newblock
  \urlprefix\url{http://www.sciencedirect.com/science/article/pii/S0959438819300248},
  \href{10.1016/j.conb.2019.08.002}{\doiprefix
  \detokenize{10.1016/j.conb.2019.08.002}}.

\bibitem[{Wilting and Priesemann(2019{\natexlab{b}})Wilting, J. and Priesemann,
  V.}]{wilting_between_2019}
\textbf{\color{eLifeMediumGrey} Wilting J}, Priesemann V.
\newblock Between {Perfectly} {Critical} and {Fully} {Irregular}: {A}
  {Reverberating} {Model} {Captures} and {Predicts} {Cortical} {Spike}
  {Propagation}.
\newblock Cerebral Cortex (New York, NY: 1991).  2019 Jun; 29(6):2759--2770.
\newblock \href{10.1093/cercor/bhz049}{\doiprefix
  \detokenize{10.1093/cercor/bhz049}}.

\bibitem[{Wilting et~al.(2018)Wilting, Jens and Dehning, Jonas and Pinheiro
  Neto, Joao and Rudelt, Lucas and Wibral, Michael and Zierenberg, Johannes and
  Priesemann, Viola}]{wilting_operating_2018}
\textbf{\color{eLifeMediumGrey} Wilting J}, Dehning J, Pinheiro~Neto J, Rudelt
  L, Wibral M, Zierenberg J, Priesemann V.
\newblock Operating in a {Reverberating} {Regime} {Enables} {Rapid} {Tuning} of
  {Network} {States} to {Task} {Requirements}.
\newblock Frontiers in Systems Neuroscience.  2018 Nov; 12.
\newblock
  \urlprefix\url{https://www.ncbi.nlm.nih.gov/pmc/articles/PMC6232511/},
  \href{10.3389/fnsys.2018.00055}{\doiprefix
  \detokenize{10.3389/fnsys.2018.00055}}.

\bibitem[{Wilting and Priesemann(2018)Wilting, Jens and Priesemann,
  Viola}]{wilting_inferring_2018}
\textbf{\color{eLifeMediumGrey} Wilting J}, Priesemann V.
\newblock Inferring collective dynamical states from widely unobserved systems.
\newblock Nature Communications.  2018 Jun; 9(1):2325.
\newblock \urlprefix\url{https://www.nature.com/articles/s41467-018-04725-4},
  \href{10.1038/s41467-018-04725-4}{\doiprefix
  \detokenize{10.1038/s41467-018-04725-4}}.

\bibitem[{Zilio et~al.(2021)Zilio, Federico and Gomez-Pilar, Javier and Cao,
  Shumei and Zhang, Jun and Zang, Di and Qi, Zengxin and Tan, Jiaxing and
  Hiromi, Tanigawa and Wu, Xuehai and Fogel, Stuart and
  others}]{zilio2021intrinsic}
\textbf{\color{eLifeMediumGrey} Zilio F}, Gomez-Pilar J, Cao S, Zhang J, Zang
  D, Qi Z, Tan J, Hiromi T, Wu X, Fogel S, et~al.
\newblock Are intrinsic neural timescales related to sensory processing?
  Evidence from abnormal behavioral states.
\newblock NeuroImage.  2021; 226:117579.

\end{thebibliography}
\end{document}